\documentclass[twocolumn]{aastex63}
\usepackage{natbib}
\usepackage{color}
\usepackage{mathtools}
\usepackage{booktabs}
\usepackage{hyperref} 

\def	\cm		{\,{\rm {cm}}}
\def	\K		{\,{\rm K}}
\def	\g		{\,{\rm {g}}}
\def	\mum	{\,{\mu \rm{m}}}

\def \bea {\begin{eqnarray}}
\def \ena {\end{eqnarray}}

\def	\bB	{\boldsymbol{B}}

\def 	\bE	{\boldsymbol{E}}

\def	\bJ	{\boldsymbol{J}}

\def	\br	{\boldsymbol{r}}

\def	\br	{{\bf r}}

\def	\cm	{\,{\rm cm}}

\def	\erg	{\,{\rm erg}}

\def	\g	{\,{\rm g}}
\def	\gas	{\,{\rm gas}}

\def	\H	{{\rm H}}

\def	\xhat		{\hat{\boldsymbol{x}}}
\def	\yhat		{\hat{\boldsymbol{y}}}
\def	\zhat		{\hat{\boldsymbol{z}}}
\def	\ahat		{\hat{\boldsymbol{a}}}

\def    \bB     	{\boldsymbol{B}} 
 
\def    \bl     	{\boldsymbol{l}} 
\def    \bE     	{\boldsymbol{E}} 
 
\def    \br     	{\boldsymbol{r}}

\newcommand{\alfven}{Alfv$\acute{\text{e}}$n~} 
\newcommand{\alfvenic}{Alfv$\acute{\text{e}}$nic~} 

\begin{document}
\shorttitle{3D B-fields using dust polarization}
\shortauthors{Hoang and Truong}
\title{Probing 3D magnetic fields using thermal dust polarization and grain alignment theory}

\author{Thiem Hoang}
\affiliation{Korea Astronomy and Space Science Institute, Daejeon 34055, Republic of Korea} 
\email{thiemhoang@kasi.re.kr}
\affiliation{Department of Astronomy and Space Science, University of Science and Technology, 217 Gajeong-ro, Yuseong-gu, Daejeon, 34113, Republic of Korea}

\author{Bao Truong}

\affiliation{Korea Astronomy and Space Science Institute, Daejeon 34055, Republic of Korea} 
\affiliation{Department of Astronomy and Space Science, University of Science and Technology, 217 Gajeong-ro, Yuseong-gu, Daejeon, 34113, Republic of Korea}


\begin{abstract}
Magnetic fields are ubiquitous in the universe and are thought to play an important role in various astrophysical processes. Polarization of thermal emission from dust grains aligned with the magnetic field is widely used to measure the two-dimensional magnetic field projected onto the plane of the sky (POS), but its component along the line of sight (LOS) is not yet constrained. Here, we introduce a new method to infer three-dimensional (3D) magnetic fields using thermal dust polarization and grain alignment physics. We first develop a physical model of thermal dust polarization using the modern grain alignment theory based on the magnetically enhanced radiative torque (MRAT) alignment theory. We then test this model with synthetic observations of magnetohydrodynamic (MHD) simulations of a filamentary cloud with our updated POLARIS code. Combining the tested physical polarization model with synthetic polarization, we show that the B-field inclination angles can be accurately constrained by the polarization degree from synthetic observations. Compared to the true 3D magnetic fields, our method based on grain alignment physics is more accurate than the previous methods that assume uniform grain alignment. This new technique paves the way for tracing 3D B-fields using thermal dust polarization and grain alignment theory and for constraining dust properties and grain alignment physics.
\end{abstract}

\section{Introduction}
Magnetic fields (B-fields) are ubiquitous in the universe and thought to play an important role in astrophysical processes, including the evolution of the interstellar medium (ISM), formation and evolution of molecular clouds and filaments, star and planet formation (see \citealt{Crutcher:2010p318, HennebelleInutsuka.2019} for reviews), and cosmic ray transport. Polarization of starlight \citep{Hall:1949p5890,Hiltner:1949p5851} and polarized thermal emission \citep{Hildebrand:1988p2573,PlanckCollaboration:2015ev} induced by aligned dust grains (hereafter dust polarization) is the most popular tool to probe the projected B-fields on the plane of the sky (POS), i.e., two-dimensional (2D) B-fields \citep{Hull:2019hw,Pattle.2022}. The polarization angle dispersion is routinely used to measure the strength of the projected B-fields, $B_{\rm POS}$, using the Davis-Chandrashekhar-Fermi (DCF) technique \citep{Davis.1951,ChandraFermi.1953}.

Significant advances in polarimetric techniques from ground-based single-dish telescopes (JCMT, \citealt{WardThompson:2017}) and interferometers, including SMA and ALMA (\citealt{Hull:2019hw}), space telescope (Planck, \citealt{PlanckCollaboration:2015ev}), and airborne telescope (SOFIA/HAWC+, \citealt{Dowell.2010}) provide a vast amount of dust polarization data, which allow us to measure the strength of 2D B-fields from a wide range of astrophysical scales, from the full galaxies to molecular clouds and filaments to prestellar cores, protostellar envelopes and disks \citep{Pattle.2022,Tsukamoto.2022s2e}. To achieve more accurate measurements of B-field strengths using dust polarization, significant efforts have been invested in improving the DCF technique over the last decades (e.g., \citealt{Ostriker.2001,Falceta.2008,Hough:2008p6066,Cho.2016,Lazarian.2022}. Yet, the measurements based on the DCF technique are still for the 2D B-fields only. Although the Zeeman splitting effect has been used to accurately measure the LOS B-fields using several spectral lines (H, OH, CN, CCS), Zeeman measurements are still seldom due to the weak signal and long-observing time \citep{Crutcher:2010p318}.

Astrophysical magnetic fields are three-dimensional, and thus accurate measurements of 3D B-fields are necessary to understand the dynamical role of B-fields in the ISM evolution and star formation \citep{Hull:2019hw,Pattle.2022}, especially in the formation and evolution of interstellar filaments \citep{HennebelleInutsuka.2019} and prestellar core collapse. However, to date, accurate measurements of interstellar 3D B-fields are not yet available. Over the last few years, various techniques have been proposed to constrain 3D B-fields by combining the different tracers, e.g., the combination of dust polarization with Faraday rotation \citep{Tahani.2018} (see \citealt{Tahani.2022} for a review), the combination of dust polarization with velocity gradient (\citealt{Lazarian.2022,HuLaz.2023,HuLaz.2023b}).

Dust polarization itself has the potential of probing 3D B-fields because the polarization angle reveals the POS component and the polarization degree constrains the inclination angle of B-fields. Indeed, \cite{Lee.1985} first presented a model for starlight polarization induced by partially aligned grains, which depends on dust properties (shape, size, and composition), degree of grain alignment with the B-field, and the inclination angle of the mean B-fields and the fluctuations of the local field with respect to the mean B-fields. Therefore, in principle, accurate measurements of the dust polarization degree can constrain the B-field's inclination angle and the tangling when dust properties and grain alignment are well constrained. Using the observed polarization degree of silicate feature at 9.7$\mum$, \cite{Lee.1985} suggested that the B-fields along the LOS toward the Backlin-Neugebauer (BN) object are inclined by an angle $\sin^{2}\gamma_{\rm max}<0.5$ and is highly ordered. Since then, there has been little progress in this direction because there was a lack of a quantitative theory of grain alignment that can accurately predict the grain alignment degree.

The last two decades have witnessed significant progress in grain alignment physics (e.g., \citealt{Hoang.2022,HoangTram.2022}, see \citealt{TramHoang.2022} for the latest reviews). The leading theory of grain alignment is based on radiative torques (RATs, \citealt{Dolginov:1976p2480,Draine:1996p6977,Lazarian:2007p2442,HoangLaz.2008}). The RAT theory is qualitatively tested with various observations \citep{Andersson.2015,LAH15}. The modern understanding of grain alignment based on RATs established that grain alignment degree depends on the fraction of grains aligned at an attractor point with angular momentum greater than the thermal value (i.e., high-J attractors), denoted by $f_{\rm high-J}$. For the RAT theory, the fraction $f_{\rm high-J}$ depends on the grain shape, the radiation field, the relative angle between the radiation direction and the magnetic field \citep{LazHoang.2007,HoangLaz.2009b}. It can span between $0.2-0.5$ \citep{Herranen.2021}. The RAT alignment theory was implemented in a polarized radiative transfer code, POLARIS \citep{Reissl.2016}, where the value of $f_{\rm high-J}$ is a fixed input parameter. The POLARIS code can predict thermal dust polarization maps and polarization spectra for different astrophysical environments.

As a first step toward an ab initio model of dust polarization, \cite{LeeHoang.2020} presented a physical model of thermal dust polarization using the RAT alignment theory, assuming the fixed $f_{\rm high-J}$ and uniform B-fields. However,
for grains with embedded iron inclusions, the value $f_{\rm high-J}$ is increased due to the joint effect of both RATs and magnetic relaxation \citep{DavisGreen.1951,Jones:1967p2924}, aka magnetically enhanced RAT alignment or MRAT \citep{HoangLaz.2008,HoangLaz.2016}. The enhanced magnetic relaxation is caused by the enhanced magnetic susceptibility of grains due to the inclusion of iron clusters into dust grains, which is most likely because Fe is among the abundant elements in the universe, and up to 95\% of Fe is missing from the gas \citep{Jenkins.2009,Dwek.2016}. For grains with embedded iron inclusions, numerical simulations reveal the increase of grain alignment efficiency with magnetic relaxation rate \citep{HoangLaz.2016,LazHoang.2019}. The MRAT alignment mechanism can now predict the degree of grain alignment as a function of the local gas density, radiation field, grain sizes, and dust magnetic properties \citep{HoangLaz.2008,HoangLaz.2016,LazHoang.2019}. Therefore, it is necessary to develop a physical model for thermal dust polarization based on the MRAT mechanism, which can be used with the observed polarization degree to infer the B-field inclination angle and then 3D B-fields.

Recently, \cite{Chen.20191om} proposed a technique to infer 3D B-fields in molecular clouds using polarization of thermal dust emission. However, the authors assumed a simple model of thermal dust polarization in which the grain alignment efficiency and the B-fields do not change within the cloud. \cite{King.2019} studied the effect of grain alignment on synthetic polarization and found it important, but their approximated alignment degree is based on the early model of RAT alignment by \cite{Cho:2005p3360} that assumed that all grains larger than a critical size for RAT alignment are perfectly aligned (i.e., ideal RAT theory).  

Moreover, the disregard of the B-field fluctuations along the LOS in \cite{Chen.20191om} is only valid for the case of strongly magnetized medium (i.e., sub-\alfven turbulence, $M_{\rm A}=\delta v/v_{\rm A}=\delta B/B<1$) where $\delta v=v_{l}$ for the turbulence velocity. However, polarimetric observations toward molecular clouds and dense clouds/cores reveal that the \alfven turbulence is usually sub-\alfvenic, $M_{\rm A}=\delta v/v_{\rm A}=\sigma_{\phi}/Q\approx 0.035\sigma_{\phi}(\textrm{degree})<0.4-1$ with $Q\sim 0.5$ (see e.g., \citealt{Pattle:2021dq,Thuong.2022,Ngoc.2023,Karoly.2023,Tram.2023}). However, the turbulence becomes super-\alfvenic with $M_{\rm A}>1$ in high-density regions of $n_{\rm H}>10^{7}\cm^{-3}$ (see Fig 2f in \citealt{Pattle.2022}). Therefore, in the majority of star-forming regions (from clouds to cores to protostellar envelope of $n_{\H}<10^{7}\cm^{-3}$, observations reveal sub-\alfvenic turbulence, i.e., the B-field fluctuations are less important due to weak \alfvenic turbulence and strong B-fields, and the polarization would depend crucially only on the inclination angle and the grain alignment efficiency. However, a different compilation of B-field strengths by \cite{Liu.2022} suggests trans-\alfvenic turbulence.

\citet{HuLaz.2022,HuLaz.2023} improved Chen's method by considering the fluctuation of the B-field along the LOS predicted by anisotropic MHD turbulence. Using the perfect alignment and MHD simulations, the authors showed that Chen's method is valid for sub-\alfvenic turbulence of $M_{\rm A}\ll 1$. Moreover, for $0.4<M_{\rm A}<1$, the deviation of Chen's result and the actual value is above 10 degrees. However, as Chen et al., \citet{HuLaz.2022,HuLaz.2023} assumed that the grain alignment degree is constant within the molecular cloud. Therefore, the variation of grain alignment must be taken into account to reliably constrain the 3D B-field based on the polarization degree. 

Here, we introduce a general technique by including the alignment efficiency using the modern theory of the MRAT grain alignment. For simplicity, we still assume the alignment efficiency is constant along the LOS, but we take into account the variation of the grain alignment efficiency across the cloud surface. By doing so, we can account for both the effect of and the grain alignment variation and magnetic turbulence across the cloud. Ultimately, one must also consider the variation of the grain alignment efficiency along the LOS, but it is challenging because of the fluctuations of both grain alignment and the magnetic turbulence.

The structure of our paper is as follows. In Section \ref{sec:method} we introduce a physical model of thermal dust polarization based on the MRAT mechanism and present a new method for constraining 3D B-fields using dust polarization. In Section \ref{sec:simul} we will test the dust polarization model using synthetic polarization observations of MHD simulations. In Section \ref{sec:results} we will present our numerical results, compare them with the polarization from the analytical model, and apply our method for inferring the inclination angles using the synthetic polarization degree map. The effects of magnetic turbulence on the dust polarization model and inclination angles are discussed in Section \ref{sec:turbulence}. An extended discussion of our results and implications is presented in Section \ref{sec:discuss}, and a summary is shown in Section \ref{sec:summ}.

\section{Methods}\label{sec:method}
\subsection{A Physical Model of Thermal Dust Polarization}
We first construct a physical model of thermal dust polarization using the theory of grain alignment based on radiative torques and magnetic relaxation \citep{HoangLaz.2016}.
\subsubsection{Grain Extinction and Polarization Cross-Section}\label{sec:align}
Consider an oblate spheroidal grain shape with the symmetry axis $\ahat_{1}$, which is also the grain axis of maximum inertia moment. The effective size of the grain of volume $V_{\rm gr}$ is $a=(3V_{\rm gr}/(4\pi))^{1/3}$. For simplicity, we consider the composite dust model in which silicate and carbonaceous material are mixed in a single dust population, which is considered the leading dust model of the ISM \citep{Draine.2021no}.

The extinction cross-sections of the spheroidal grain for the electric field of the incident light parallel/perpendicular to the symmetry axis are denoted by $C_{\|}=C(\bE\| \ahat_{1}), C_{\perp}=C(\bE\perp \ahat_{1})$, respectively. The polarization cross-section of the oblate grain is given by $C_{\rm pol}=C_{\perp}-C_{\|}$. The total extinction cross-section, obtained by averaging the extinction cross-section over an ensemble of grains with random orientation, is approximately given by $C_{\rm ext}=(2C_{\perp}+C_{\|})/3$ (see e.g., \citealt{Hoang.2013,Draine.2021no}).

Let $dn/da$ be the grain size distribution. Then, $n_{d}(a)=n_{\rm H}^{-1}dn/da$ with $n_{\rm H}$ being the hydrogen number density is the density of grains with a size in the range of $a,a+da$ per H. Here, we assume the power-law size distribution of $n_{\rm H}^{-1}dn/da=Ca^{-3.5}$ with $C$ the normalization constant \citep{Mathis.1977}.

The dust extinction and polarization coefficients are calculated as
\bea
\sigma_{\rm ext}(\lambda)=\int_{a_{\rm min}}^{a_{\rm max}}da n_{d}(a)C_{\rm ext}(a,\lambda),\\
\sigma_{\rm pol}(\lambda)=\int_{a_{\rm min}}^{a_{\rm max}}da n_{d}(a)C_{\rm pol}(a,\lambda),\label{eq:extpol_coeff}
\ena
where $a_{\rm min}$ and $a_{\rm max}$ are the minimum and maximum grain sizes, respectively.

\subsubsection{Thermal Dust Polarization from Aligned Grains}
\begin{figure}
    \includegraphics[width=0.5\textwidth]{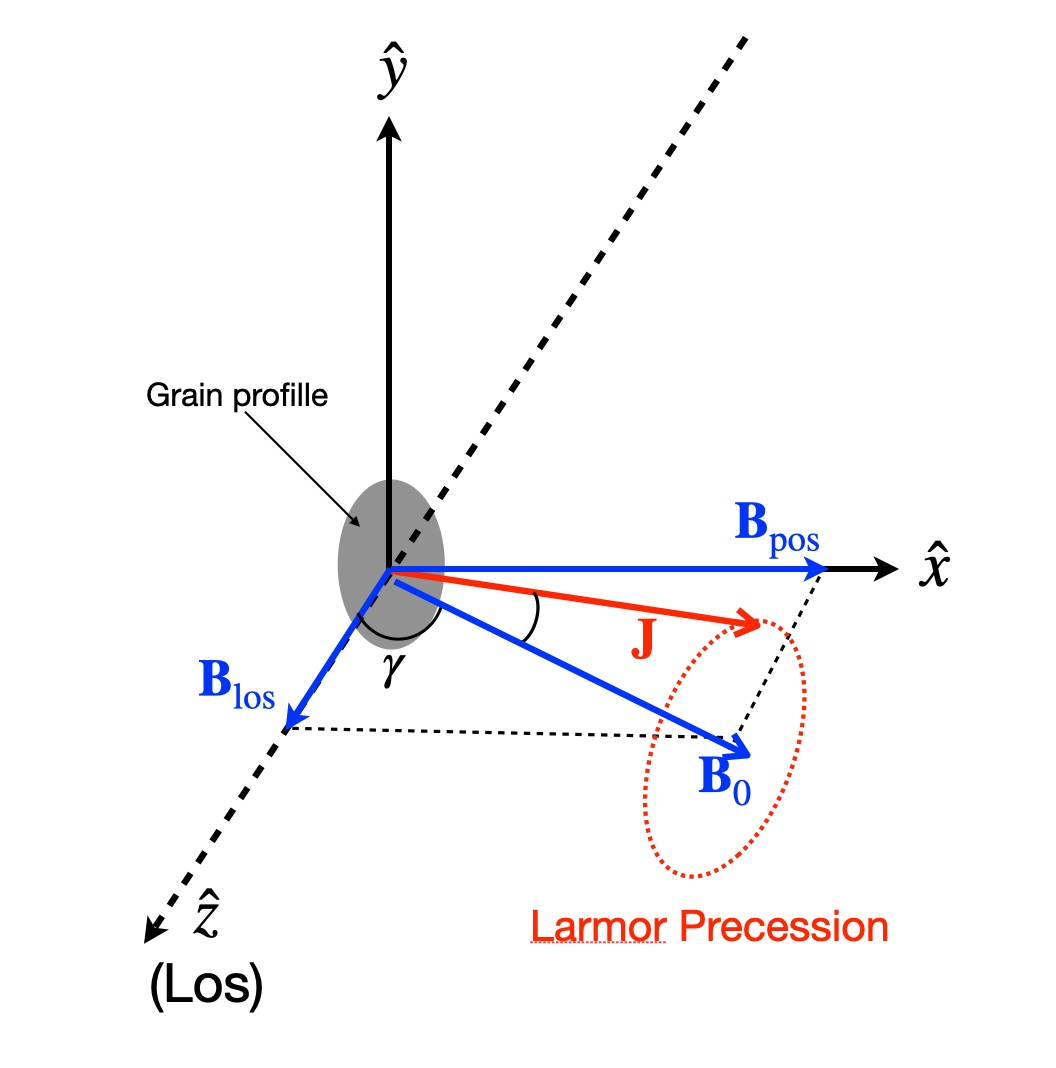}
    \caption{Illustration of grain alignment with the mean B-field $\bB$, the line of sight, $\zhat$, and the sky plane $\xhat\yhat$. The mean B-field lies in the plane $\xhat \zhat$ and makes an angle $\gamma$ with the line of sight.}
    \label{fig:JB_Cxy}
\end{figure}

The modern grain alignment theory establishes that interstellar grains are aligned with their axis of maximum inertia moment, also the grain's shortest axis ($\ahat_{1}$) along the angular momentum $\bJ$ due to various internal relaxation processes, including Barnett relaxation \citep{Purcell.1979}, inelastic relaxation \citep{LazEfroim.1999,Efroimsky:2000p5384} and nuclear relaxation \citep{LazDraine.1999}. The grain angular momentum $\bJ$ rapidly precesses around the ambient B-field $\bB$ due to fast Larmor precession which determines the axis of grain alignment in the majority of astrophysical environments (see \citealt{Hoang.2022} for details). Let $\xi$ be the alignment angle between $\bJ$ and $\bB$ (see Figure \ref{fig:JB_Cxy}). The net alignment degree of the grain's axis of maximum inertia moment with $\bB$ is described by $f_{\rm align}(a)$.

Let $\zhat$ be the line of sight (propagation direction of light). The observer's coordinate system is described by $\xhat\yhat\zhat$ with $\xhat\yhat$ representing the POS. For convenience, we assume that the mean B-field $\bB_{0}$ lies in the plane $\xhat\zhat$ and makes an inclination angle $\gamma$ with respect to $\zhat$, and the component projected onto the POS, $B_{\rm pos}$, is directed along the $\xhat$ axis (see Figure \ref{fig:JB_Cxy}). Let $C_{x}, C_{y}$ be the extinction cross-sections for the incident electric field $\bE$ along $\xhat$ and $\yhat$, respectively.

The polarization and extinction cross-section averaged over the grain's Larmor precession for the light propagating along the $\zhat$ are given by ( \citealt{Lee.1985,Hoang.2013,PlanckXX.2015})
\bea
\frac{C_{y}-C_{x}}{2}&=&C_{\rm pol}f_{\rm align}\sin^{2}\gamma,\\
\frac{C_{y}+C_{x}}{2}&=&C_{\rm ext}+C_{\rm pol}f_{\rm align}(2/3-\sin^{2}\gamma).\label{eq:Cpol_xy}
\ena
 
For the sake of simplicity, we first assume that grain alignment, dust properties, and B-fields do not change along the LOS. As a result, within the modified picket-fence approximation (MPFA) model of polarization \citep{Dyck.1974,Draine.2021no}, the intensity of polarized and total thermal emission for a composite dust model of mixed silicate and carbonaceous material is 
\bea
\frac{I_{\rm pol}(\lambda)}{N_{\rm H}B_{\lambda}(T_{d})}&=&\int_{a_{\rm min}}^{a_{\rm max}}da n_{d}(a)C_{\rm pol}f_{\rm align}(a)\sin^{2}\gamma, \\
\frac{I_{\rm em}(\lambda)}{N_{\rm H}B_{\lambda}(T_{d})}&=&\int_{a_{\rm min}}^{a_{\rm max}}da n_{d}(a)[C_{\rm ext}+C_{\rm pol}f_{\rm align}(2/3-\sin^{2}\gamma)], 
\ena
where $T_{d}$ is the mean dust temperature\footnote{Here, for the physical model, we ignore the dependence of the grain temperature on the grain size and assume all grain sizes have the same mean temperature.}, and $N_{\rm H}$ is the hydrogen gas column density along the LOS (see also Appendix \ref{sec:apdx}). 

In the electric dipole limit of $\lambda/2\pi a\gg 1$, the cross-section is proportional to the grain volume, i.e., $C_{\rm pol,ext}\propto V_{\rm gr}(a)=4\pi a^{3}/3$. Let $C_{\rm pol}(a)=\eta_{\rm pol} V_{\rm gr}(a)$, then, the polarized intensity can be written as
\bea
\frac{I_{\rm pol}(\lambda)}{N_{\rm H}B_{\lambda}(T_{d})}&=&\eta_{\rm pol}\int_{a_{\rm min}}^{a_{\rm max}}da n_{d}(a)(4\pi a^{3}/3)f_{\rm align}\sin^{2}\gamma,\\
\frac{I_{\rm pol}(\lambda)}{N_{\rm H}B_{\lambda}(T_{d})}&=&\eta_{\rm pol}V_{d}\langle f_{\rm align}\rangle\sin^{2}\gamma=\sigma_{\rm pol}\langle f_{\rm align}\rangle\sin^{2}\gamma,\label{eq:Ipol_avg}
\ena
where the mass-weighted alignment efficiency is given by
\bea
\langle f_{\rm align}\rangle = \frac{\int_{a_{\rm min}}^{a_{\rm max}}da (4\pi a^{3}/3)n_{d}(a)f_{\rm align}}{V_{d}},\label{eq:falign_mass}
\ena
and $V_{d}=\int (4\pi a^{3}/3)n_{d}(a)da$ is the total dust volume per H respectively, 
and $\eta_{\rm pol}=\sigma_{\rm pol}/V_{d}$. Similarly, the emission intensity can be rewritten as
\bea
\frac{I_{\rm em}(\lambda)}{N_{\rm H}B_{\lambda}(T_{d})}=\sigma_{\rm ext}+\sigma_{\rm pol}\langle f_{\rm align}\rangle (2/3-\sin^{2}\gamma). \label{eq:Iem_avg}
\ena

From Equations (\ref{eq:Ipol_avg}) and (\ref{eq:Iem_avg}), we calculate the degree of thermal dust polarization:
\bea
p(\lambda)&=&\frac{I_{\rm pol}}{I_{\rm em}}=\frac{\sigma_{\rm pol} \langle f_{\rm align}\rangle\sin^{2}\gamma}{\sigma_{\rm ext}-\langle f_{\rm align}\rangle(\sin^{2}\gamma-2/3)\sigma_{\rm pol}},\nonumber\\
&=&\frac{p_{i} \langle f_{\rm align}\rangle\sin^{2}\gamma}{1-p_{i}\langle f_{\rm align}\rangle(\sin^{2}\gamma-2/3)},\label{eq:pol}
\ena
where the extinction and polarization coefficients are given by Equations (\ref{eq:extpol_coeff}), and 
$p_{i}= \sigma_{\rm pol}/\sigma_{\rm ext}$ is the intrinsic polarization degree which only depends on the grain properties (shape and size distribution), i.e., larger $p_{i}$ for larger elongation (e.g., \citealt{Lee.1985,Draine.2021no}). 

Equation (\ref{eq:pol}) reveals that the dust polarization degree depends on three key parameters, including intrinsic polarization ($p_{i}$, i.e., the grain shape), grain alignment degree ($\langle f_{\rm align}\rangle)$, and the projection of the mean B-field onto the POS ($\sin^{2}\gamma$).

\subsubsection{Grain Alignment Theory}
\label{sec:alignment_theory}
The key parameter required for the polarization model (Eq. \ref{eq:pol}) is the grain alignment function, $f_{\rm align}(a)$. Here, we derive $f_{\rm align}$ using the modern theory of grain alignment based on radiative torques (RATs) and magnetic relaxation.

Traditionally, the alignment efficiency for an ensemble of grains with size $a$ is described by the Rayleigh reduction factor, $R$ \citep{Greenberg:1968p6020}. The Rayleigh reduction factor describes the average alignment degree of the axis of major inertia of grains ($\ahat_{1}$) with its angular momentum ($\bJ$, i.e., internal alignment) and of the angular momentum with the ambient B-field (i.e., external alignment), given by
\bea
R=\langle \frac{1}{2}(3\cos^{2}\beta-1)\frac{1}{2}(3\cos^{2}\xi-1)\rangle,\label{eq:Ralign} 
\ena
where $\beta$ is the angle between $\ahat_{1}$ and $\bJ$ and $\xi$ is the angle between $\bJ$ and $\bB$, and $\langle...\rangle$ describes the averaging over an ensemble of grains (see, e.g., \citealt{Hoang.2014,HoangLaz.2016,HoangLaz.2016b}). Here, $R=0$ for randomly oriented grains, and $R=1$ for perfect internal and external alignment.

Therefore, we can define the grain alignment function as follows \citep{Hoang.2014}:
\bea
f_{\rm align}(a)=R.\label{eq:falign}
\ena

$\bullet$ {\bf Grain alignment by RATs only}\\
In general, grain alignment by RATs can occur at low-J attractors and high-J attractors \citep{Weingartner.2003,HoangLaz.2008,HoangLaz.2016}. Let $f_{\rm high-J}$ be the fraction of grains that can be aligned by RATs at high-J attractors. For grain alignment by only RATs (e.g., grains of ordinary paramagnetic material, \citealt{DavisGreen.1951}), high-J attractors are only present for a limited range of the radiation direction that depends on the grain shape \citep{DraineWein.1997,HoangLaz.2008,Herranen.2019}.

Numerical simulations in \cite{HoangLaz.2008,HoangLaz.2016} show that if the RAT alignment has a high-J attractor point, then, large grains can be perfectly aligned because grains at low-J attractors would be randomized by gas collisions and eventually transported to more stable high-J attractors by RATs. On the other hand, grain shapes with low-J attractors would have negligible alignment due to gas randomization. For small grains, numerical simulations show that the alignment degree is rather small even in the presence of iron inclusions because grains rotate subthermally \citep{HoangLaz.2016}.

The first parameter of the RAT alignment theory is the minimum grain size required for grain alignment, denoted by $a_{\rm align}$. Let $n_{\H}$ and $T_{\gas}$ be the local gas density and temperature. Let $\gamma_{\rm rad}$, $u_{\rm rad}$, and $\bar{\lambda}$ be the anisotropy degree, radiation energy density, and the mean wavelength of the radiation field. The minimum alignment size by RATs is given by \cite{Hoang.2021}:
\bea
a_{\rm align}&=&\left(\frac{1.2n_{\rm H}T_{\rm gas}}{\gamma_{\rm rad} u_{\rm rad}\bar{\lambda}^{-2}} \right)^{2/7}\left(\frac{15m_{\rm H}k^{2}}{4\rho}\right)^{1/7}(1+F_{\rm IR})^{2/7}\nonumber\\
&\simeq &0.055\hat{\rho}^{-1/7} \left(\frac{\gamma_{-1}U}{n_{3}T_{\rm gas,1}}\right)^{-2/7} \nonumber\\
    &&\times \left(\frac{\bar{\lambda}}{1.2\mum}\right)^{4/7} (1+F_{\rm IR})^{2/7} ~\mum,\label{eq:aalign_ana}
\ena 
where $\hat{\rho}=\rho/(3\g\cm^{-3})$ is the normalized mass density of grain material, $T_{\rm gas,1}=T_{\rm gas}/10\K$, $n_{3}=n_{\H}/(10^{3}\cm^{-3})$, $\gamma_{-1}=\gamma_{\rm rad}/0.1$, $U=u_{\rm rad}/u_{\rm ISRF}$ with $u_{\rm ISRF}=8.6\times 10^{-13}\erg\cm^{-3}$ being the radiation energy density of the interstellar radiation field (ISRF) in the solar neighborhood \citep{Mathis.1983}, and $F_{\rm IR}$ is a dimensionless parameter that describes the grain rotational damping by infrared emission. For the ISM and dense clouds, $F_{\rm IR}\ll 1$, and can be omitted in Equation (\ref{eq:aalign_ana}).

Therefore, the Rayleigh reduction factor can be parameterized by the fraction of grains aligned at low-J and high-J attractors \citep{Hoang.2014}:
\bea
R=
\left\{
\begin{array}{l l}   
0 ~  {\rm ~ for~ } a < a_{\rm align}\\
f_{\rm high-J}Q_{X}^{\rm high-J}+ (1-f_{\rm high-J})Q_{X}^{\rm low-J}~  {\rm ~ for~}  a\gtrsim  a_{\rm align}\\
\end{array}\right..
\label{eq:Ralign}
\ena
where $Q_{X}=\langle (3\cos^{2}\beta-1)/2\rangle
$ describes the internal alignment of grain axes with $\bJ$ \citep{RobergeLaz.1999}, and the external alignment of $\bJ$ with $\bB$ is assumed to be perfect \citep{HoangLaz.2008,HoangLaz.2016}. At high-J attractors, $Q_{X}^{\rm high-J}$ can reach $100\%$ due to suprathermal rotation, but at low-J attractors with thermal rotation, the value of $Q_{X}^{\rm low-J}$ is smaller which follows the local thermal equilibrium (Boltzmann) distribution of the angle $\beta$ with roughly $30\%$ \citep{HoangLaz.2016,HoangLaz.2016b}.\footnote{The calculations of $Q_{X}$ using the Boltzmann distribution are valid for grains with fast internal relaxation only, which is appropriate for our study here because grains are small and have magnetic inclusions. For very large grains, internal relaxation can be rather slow \citep{Hoang.2022}.} Therefore, the second term is subdominant, and the alignment efficiency $R$ only depends on $f_{\rm high-J}$. 

$\bullet$ {\bf Grain alignment by MRAT}\\
Composite grains considered in this paper contain silicate, carbonaceous material, and clusters of metallic iron or iron oxide. Let $N_{\rm cl}$ be the number of iron atoms per cluster, and $\phi_{\rm sp}$ be the volume filling factor. The existence of metallic iron/iron oxide clusters makes composite grains become superparamagnetic material that has magnetic susceptibility increased by a factor of $N_{\rm cl}$ from ordinary paramagnetic material \citep{DavisGreen.1951,Jones:1967p2924}. The enhanced magnetic susceptibility significantly increases the rate of magnetic relaxation (denoted by $\tau_{\rm mag,sp}^{-1}$) and the degree of grain alignment (\citealt{LazHoang.2008,HoangLaz.2016,LazHoang.2019}). 

The strength of magnetic relaxation for rotating composite grains is defined by the ratio of  the magnetic relaxation rate relative to the gas randomization rate (denoted by $\tau_{\rm gas}^{-1}$)
\bea
\delta_{\rm mag}&=&\frac{\tau_{\rm mag,sp}^{-1}}{\tau_{\rm gas}^{-1}}\nonumber\\
&= &5.6a_{-5}^{-1}\frac{N_{\rm cl}\phi_{\rm sp,-2}\hat{p}^{2}B_{2}^{2}}{\hat{\rho} n_{3}T_{\gas,1}^{1/2}}\frac{k_{\rm sp}(\Omega)}{T_{d,1}},\label{eq:delta_m}~~~~
\ena
where $T_{\rm d,1}=T_{\rm d}/10\K$ is the normalized dust temperature, $\hat{p}=p/5.5$ where $p\mu_{B}$ is the mean magnetic moment per iron atom and $\mu_{B}=e\hbar/2m_{e}c$ is the Bohr magneton, $B_{2}=B/10^{2}\mu G$ is the normalized B-field strength, and $\phi_{sp,-2}=\phi_{\rm sp}/10^{-2}$. Above, $k_{\rm sp}(\Omega)$ is the function of the grain angular velocity $\Omega$ which describes the suppression of the magnetic susceptibility at high angular velocity (see, e.g., \citealt{Hoang.2022}).

Detailed numerical calculations in \cite{HoangLaz.2016} show the increase of $f_{\rm high-J}$ and $R$ with $\delta_{\rm mag}$. To model the increase of $f_{\rm high-J}$ with $\delta_{m}$, as in \cite{Giang.2023}, we introduced the following parametric model:
\bea 
f_{\rm high-J}(\delta_{\rm mag}) = \left\{
\begin{array}{l l}    
    0.25 ~ ~  {\rm ~ for~ } \delta_{\rm mag} < 1   \\
    0.5 ~ ~  {\rm ~ for~ } 1 \leq \delta_{\rm mag} \leq 10 \\
    1    ~ ~ ~ ~  {\rm ~ for~}  \delta_{\rm mag} > 10 \\
\end{array}\right..
\label{eq:fhiJ_deltam}
\ena
Above, for alignment grain driven by RATs only, i.e., for $\delta_{\rm mag}<1$, we assumed $f_{\rm high-J}=0.25$. This choice is based on a statistical study of grain alignment by RATs for an ensemble of grain shapes in \cite{Herranen.2019}.

For a given grain size distribution and magnetic properties, using $a_{\rm align}$, $\delta_{\rm mag}$ and $R$, one can calculate $\langle f_{\rm align}\rangle$ by plugging $f_{\rm align}(a)$ into Equation (\ref{eq:falign_mass}). It can be seen that $\langle f_{\rm align}\rangle$ cannot reach $1$ because there exists a fraction of small dust of size below $a_{\rm align}$ which are weakly not aligned. Therefore, even in the ideal RAT theory, the maximum value of $\langle f_{\rm align}\rangle<1$.

\subsection{Constraining the B-field inclination angle}
Equation (\ref{eq:pol}) shows the dependence of the dust polarization degree on three parameters, including the intrinsic polarization degree, $p_{i}$, the mass-averaged alignment efficiency, $\langle f_{\rm align}\rangle$, and the inclination angle $\gamma$. For a given grain shape, the intrinsic polarization is known \citep{Draine.2021no}. The MRAT alignment theory yields $\langle f_{\rm align}\rangle$. Therefore, if the thermal dust polarization along a LOS is measured to a degree $p_{\rm obs}$, we can infer the mean inclination angle of the B-field along this LOS by setting the polarization degree predicted from Equation (\ref{eq:pol}) to $p_{\rm obs}$, as follows:
\bea
\sin^{2}\gamma = \frac{p_{\rm obs}(1+2p_{i}\langle f_{\rm align}\rangle/3)}{p_{i}\langle f_{\rm align}\rangle(1+p_{\rm obs})}.\label{eq:chi2_align}
\ena

Equation (\ref{eq:chi2_align}) provides the direct link between the inclination angle as a function of the observed polarization degree, intrinsic polarization, and grain alignment. Therefore, when the latter parameters are constrained, one can infer the inclination angle.

\subsection{The idealized case of uniform grain alignment}
\cite{Chen.20191om} considered the idealized case of uniform grain alignment and assumed the constant polarization, $p_{0}=p_{i}\langle f_{\rm align}\rangle$. In this special case, the polarization degree (Eq. \ref{eq:pol}) and the inclination angle (Eq. \ref{eq:chi2_align}) return to the results in \cite{Chen.20191om}. The latter is given by
\bea
(\sin^{2}\gamma)_{\rm Chen} = \frac{p_{\rm obs}(1+2p_{0}/3)}{p_{0}(1+p_{\rm obs})},\label{eq:chi2_PA}
\ena
where the subscript ${\rm Chen}$ denotes the results from \cite{Chen.20191om}.

\cite{Chen.20191om} obtained the polarization $p_{0}$ using the maximum polarization observed toward the cloud, $p_{\rm max}$, which requires the B-field lying in the POS (i.e., $\sin^{2}\gamma=1$) and perfect grain alignment of $f_{\rm align}=1$. Using these criteria for Equation (\ref{eq:pol}) one obtains
\bea
p_{\rm max}=\frac{3p_{0}}{3-p_{0}},\label{eq:pmax}
\ena
which yields,
\bea
p_{0}=\frac{3p_{\rm max}}{3+p_{\rm max}}.\label{eq:p0}
\ena

The determination of the polarization parameter, $p_{0}$, from Equation (\ref{eq:p0}), in practicality, is uncertain because the B-fields within the observed region may not lie in the POS or the grain alignment is not perfect, which does not satisfy the criteria for the maximum polarization. In our proposed method, the intrinsic polarization, $p_{i}$, is based on the input grain elongation, which can be constrained with polarization observations (see e.g., \citealt{Draine.2021no}).

Comparing Equation (\ref{eq:chi2_align}) with (\ref{eq:chi2_PA}), one can see that the difference between the inclination angle for the realistic case with alignment is different from that of perfect alignment by
\bea
\sin\gamma=\frac{(\sin\gamma)_{\rm PA}}{\langle f_{\rm align}\rangle^{1/2}},\label{eq:sin_gamma_PA}
\ena
where we have assumed $1+p_{\rm obs}\approx 1$ and $1+2p_{0}\langle f_{\rm align}\rangle \approx 1$.

Therefore, the alignment degree affects significantly the inclination angle inferred from the perfect alignment (PA) model. Table \ref{tab:angle_align} shows the importance of grain alignment. The variation of the inclination angles in the presence of grain alignment, assuming a fixed maximum polarization and observed polarization degree of $p_{\rm obs}=p_{\rm max}/2$. The perfect alignment yields the inclination angle of $41.25^{\circ}$, but the imperfect alignment yields a wide range of the inclination angle. If the alignment efficiency is decreased to $60\%$, the inclination angle is decreased by a factor of 2. The inclination angle cannot be retrieved if the alignment degree is reduced to less than $50\%$.

\begin{table}[]
    \caption{Effects of grain alignment on inferred inclination angles.}
    \begin{tabular}{l l l l l}
    \toprule
      $p_{\rm max}$   &  $p_{\rm obs}$ & $\langle f_{\rm align}\rangle$ & $\sin^{2}\gamma$ & $\gamma(degree)$ \\
      \midrule
      0.3   & 0.15 & 1 & 0.565 & 41.25\cr
      0.3   & 0.15 & 0.9 & 0.618 & 38.15\cr
      0.3   & 0.15 & 0.8 & 0.684 & 34.15\cr
      0.3   & 0.15 & 0.7 & 0.770 & 28.64\cr
      0.3   & 0.15 & 0.6 & 0.884 & 19.90\cr
      0.3   & 0.15 & 0.5 & 1.04 & NA\cr
      0.3   & 0.15 & 0.4 & 1.28 & NA\cr
      \hline
      0.3   & 0.05 & 1 & 0.20 & 62.98\cr
      0.3   & 0.05 & 0.9 & 0.22 & 61.63\cr
      0.3   & 0.05 & 0.8 & 0.25 & 59.99\cr
      0.3   & 0.05 & 0.7 & 0.28 & 57.97\cr
      0.3   & 0.05 & 0.6 & 0.32 & 55.38\cr
      0.3   & 0.05 & 0.5 & 0.38 & 51.88\cr
      0.3   & 0.05 & 0.4 & 0.46 & 46.82\cr
      0.3   & 0.05 & 0.3 & 0.61 & 38.42\cr
      0.3   & 0.05 & 0.2 & 0.90 & 17.97\cr
      \bottomrule
    \end{tabular}
    \label{tab:angle_align}
\end{table}

Equation (\ref{eq:chi2_align}) has the physical solution of $\gamma$ only when $\sin^{2}\gamma\lesssim 1$. However, in reality, one may encounter singular situations of $\sin^{2}\gamma>1$ because the alignment efficiency $f_{\rm align}\ll 1$, as shown in Table \ref{tab:angle_align}.

\section{Synthetic polarization observations of MHD simulations}\label{sec:simul}
The physical model of thermal dust polarization described by Equation (\ref{eq:pol}) is obtained assuming the homogeneous gas, dust, and B-field properties along the LOS. In reality, these properties change along the LOS due to the turbulent nature of the ISM and grain evolution. Therefore, to infer the inclination angle using the observed polarization degree, we first need to test whether the physical polarization model can adequately reproduce the observed polarization. To do so, we will post-process MHD simulations with our POLARIS code to obtain the synthetic dust polarization and compare it with the results obtained from the model. 

\subsection{MHD simulation datacube}
We use a snapshot taken at time $t = 0.783$ kyr of MHD simulations that follows the evolution of a filamentary molecular cloud under the effects of self-gravitating and B-fields by \cite{Ntormousi.2019}. The filamentary cloud was set up with a length of 66 pc and an effective thickness of 33 pc onto a three-dimensional data cube of $256^{3}$ pixels. The cloud is supersonic with $M_{\rm s} \sim 5 - 10$, and the gas motion is mostly driven by self-gravity and turbulent fragmentation. The turbulence is initially added with the initial ratio of the free-fall time and the turbulence crossing time $q_{\rm turb} = t_{\rm ff}/t_{\rm turb}$ set to unity, and the energy spectrum followed by Kolmogorov’s power law with the slope of -5/3. The initial mean B-field has a strength of $B=5\mu$G and is perpendicular to the filament axis. The simulation box size of 66 pc was chosen to cover the whole morphology of the cloud.  

\begin{figure*}
    \centering
    \includegraphics[width = 0.48\textwidth]{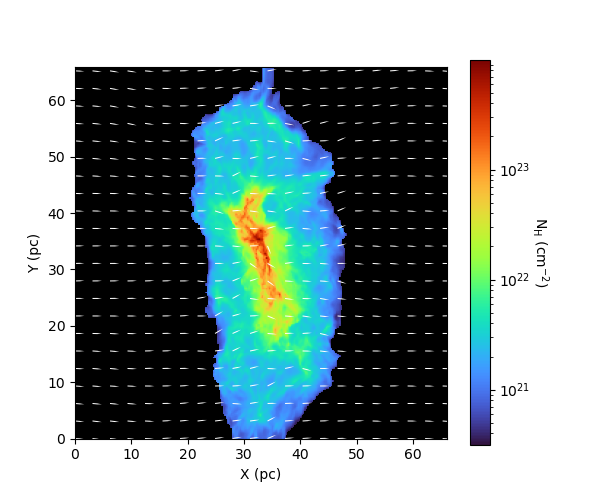}
    \includegraphics[width = 0.48\textwidth]{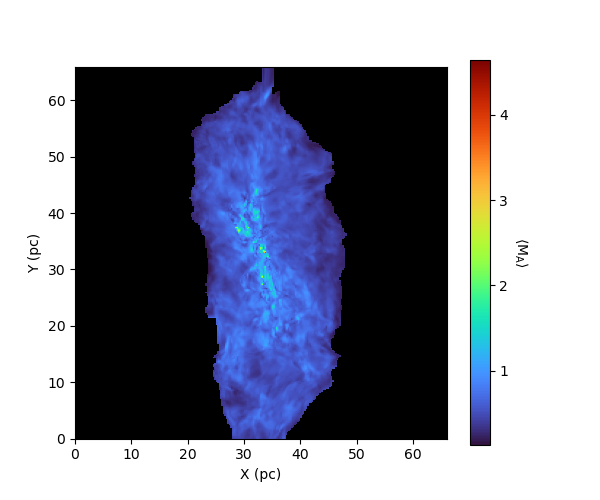}
    \caption{Map of the gas column density (left panel) and the \alfven Mach number (right panel) from MHD simulations of the filamentary cloud. The cloud is mostly sub-\alfvenic, with the mean geometry of the B-field along the x-axis of the plane-of-sky, illustrated by the white segments (left panel). The gas density varies considerably within the cloud, which is expected to affect the grain alignment efficiency and the inferred inclination angles through dust polarization.}
    \label{fig:MHD_data}
\end{figure*}

Figure \ref{fig:MHD_data} shows the maps of the gas column density with the segments showing the B-field orientation (left panel) and the \alfven Mach number averaged along the LOS (right panel) from the MHD datacube. The gas column density changes significantly within the filament, from $N_{\rm H} \sim 10^{21}-10^{22}\,\cm^{-2}$ in the outer region to $N_{\rm H} \sim 4-5\times 10^{23}\,\cm^{-2}$ in the central region. The filament is mostly sub-\alfvenic with $\langle M_{\rm A} \rangle \sim 0.2 - 0.8$.



\subsection{Dust and Grain Alignment Models}
Here, we consider the composite model of interstellar dust with $67.5\%$ of silicate and $32.5\%$ of carbonaceous grains. The choice of composite dust model is advantageous because the polarization degree at long wavelengths in Equation (\ref{eq:pol}) is independent of dust temperature. It is also motivated by the latest studies that advocate for a single dust component of the ISM \citep{Draine.2021b1e,Hensley.2023}. 
The grain size distribution is described by the power law with $dn \varpropto Ca^{-3.5}da$ in the range of grain sizes from $a_{\rm min} = 5\,\rm nm$ to $a_{\rm max} = 0.25\,\rm\mu m$ \citep{Mathis.1977}. The grain shape is assumed to be oblate spheroid with the axial ratio of $s = 2$. The choice of such an elongation is for the sake of convenience because of the existing pre-calculated cross-sections in the POLARIS code. However, it only affects the intrinsic polarization $p_{i}$ and does not affect the resulting conclusions (see Appendix \ref{sec:astrodust} for the case of smaller elongation with $s=1.4$).

To quantify the effect of grain alignment on synthetic dust polarization and inferred inclination angle, we consider the different grain alignment models. Table \ref{tab:align_model} shows the list of considered grain alignment models, including perfect alignment (PA), ideal RAT, and MRAT with different magnetic properties. For the radiation field, we consider grains to be irradiated by only the ISRF, which is described by the average radiation field in the solar neighborhood \citep{Mathis.1977}. For grain magnetic properties, we consider both paramagnetic grains (PM) with the fraction of iron of $f_p = 0.1$, and superparamagnetic (SPM) grains with increasing levels of iron inclusion $N_{\rm cl}$ from $50$ to $10^3$. Following the MRAT mechanism \citep{HoangLaz.2016}, SPM grains are predicted to achieve a higher alignment degree than PM grains due to higher magnetic susceptibility, with a higher fraction of grains being aligned with B-fields (see Equation \ref{eq:falign} - \ref{eq:fhiJ_deltam}).

\begin{table}[]
    \caption{Grain alignment models considered.}
    \begin{tabular}{l l l}
    \toprule
      Model  &  Aligned Sizes & Parameters \\
      \midrule
      PA   &  $[a_{\rm min},a_{\rm max}]$& $R=1$ \cr
      Ideal RAT  &$[a_{\rm align},a_{\rm max}]$  & $R=1$ \cr
      MRAT   & $[a_{\rm align},a_{\rm max}]$ & PM, $f_{p}=0.1$ \cr
          & $[a_{\rm align},a_{\rm max}]$ &  SPM, $N_{\rm cl}=50,10^{2},10^{3}$ \cr
      \bottomrule
    \end{tabular}
    \label{tab:align_model}
\end{table}

\subsection{Synthetic Polarization Modeling with POLARIS}
The RAT alignment physics was implemented in the POLARIS code by \cite{Reissl.2016}, whereas the MRAT mechanism was recently incorporated in the updated version by \cite{Giang.2023}. Here, we use the latest version of POLARIS code by \cite{Giang.2023} for our numerical studies. For a detailed description of the implementation of RAT alignment physics and polarized radiative transfer, please see \cite{Reissl.2016}. For a detailed description of the alignment model parameters, see \cite{Giang.2023}.  

\section{Numerical Results}\label{sec:results}
Here we first show the numerical results from synthetic polarization observations and compare them with the polarization calculated by the analytical model. We consider the different viewing angles, denoted by $\gamma_{\rm view}$, which is defined be the angle between the mean B-field $\bB_{0}$ and the LOS (see Figure \ref{fig:JB_Cxy}). We will then apply our technique to infer the inclination angle of the B-field using the synthetic data and obtain the full 3D B-field strength. 

\subsection{Grain alignment size and mass-weighted grain alignment efficiency}

For numerical calculations of the mass-weighted alignment degree with POLARIS, denoted by $\langle f_{\rm align}\rangle$, we calculate $f_{\rm align}$ for each cell using the alignment size $a_{\rm align}$ and $\delta_{\rm mag}$ and integrate it along the LOS. 

\begin{figure*}
    \centering
    \includegraphics[width = 1\textwidth]{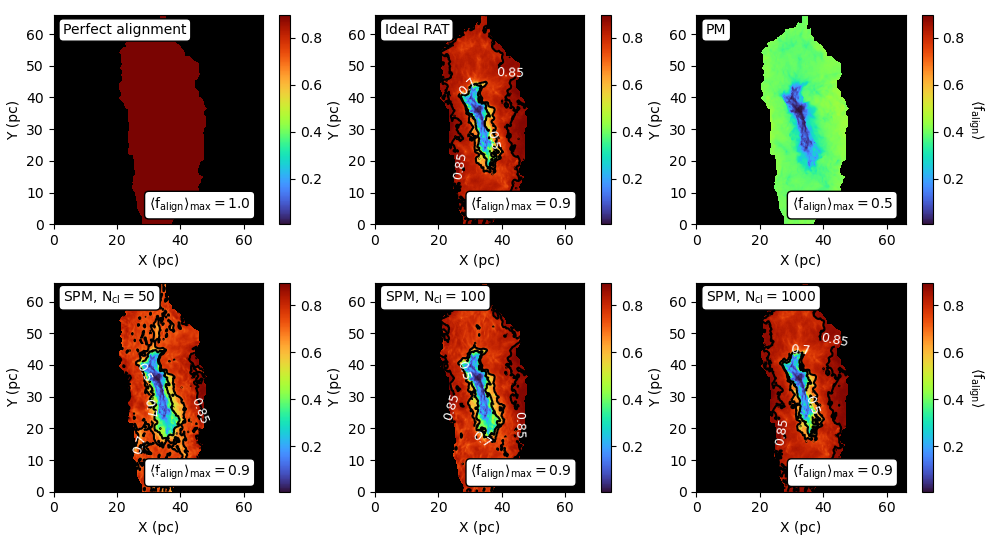}
    \caption{Map of mass-weighted grain alignment efficiency from POLARIS for the different alignment models. The black contours show the grain alignment efficiency of 0.5, 0.7, and 0.85, respectively. The alignment efficiency is uniform with $\langle f_{\rm align} \rangle = 1$ for the perfect alignment case. However, the grain alignment is strongly affected by the gas randomization, resulting in the variation of alignment efficiency within the cloud with $\langle f_{\rm align} \rangle < 1$ in the Ideal RAT case. The alignment efficiency is much lower to $\langle f_{\rm align} \rangle < 0.5$ when grains are paramagnetic, and can be enhanced for SPM grains having high levels of iron inclusions.}
    \label{fig:falign_map}
\end{figure*}

Figure \ref{fig:falign_map} shows the map of the mass-weighted grain alignment efficiency for the different alignment models, including perfect alignment, ideal RAT alignment, and MRAT alignment (see Table \ref{tab:align_model} for details), assuming the mean B-field in the POS ($\gamma_{\rm view}=90^{\circ}$). For the model of perfect alignment, all grain sizes are perfectly aligned with B-fields with a constant $\langle f_{\rm align} \rangle = 1$ in the entire filament. In the case of ideal RAT, by contrast, the alignment degree is strongly affected by gas randomization, particularly in the inner regions with high gas density (see, e.g., \citep{Hoang.2021}). Then, the value of $\langle f_{\rm align} \rangle$ decreases from 0.9 to 0.1 along the filamentary cloud. 

For the models of MRAT alignment, the grain alignment efficiency significantly depends on the magnetic properties of grains. When grains are paramagnetic, only $25\%$ of their population can be at high-J attractors by RATs, while the remaining grains at low-J attractors are aligned with a lower alignment degree of $30\%$ (see Section \ref{sec:alignment_theory}). This produces a lower alignment efficiency with $\langle f_{\rm align} \rangle < 0.5$. For SPM grains, the increasing levels of embedded iron enhance the magnetic relaxation and the fraction of grains at high-J attractors (see Equation \ref{eq:fhiJ_deltam}). Consequently, these grains can be efficiently aligned by MRAT, with a maximum alignment efficiency up to $\langle f_{\rm align}\rangle=0.9$, the same as the maximum value achieved for the Ideal RAT model.

\begin{figure}
    \includegraphics[width = 0.48\textwidth]{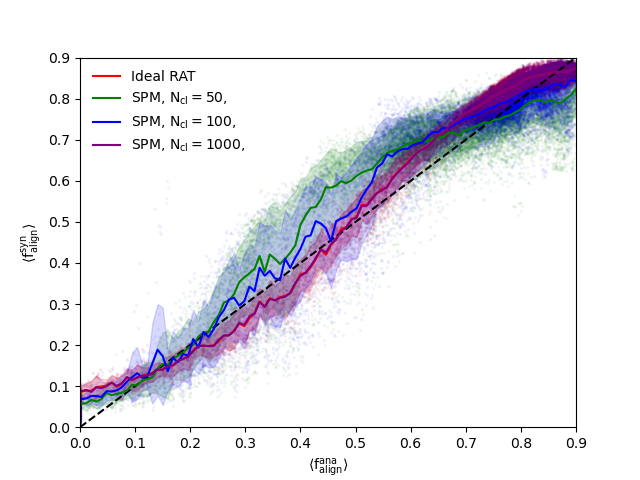}
    \caption{Comparison of mass-weighted alignment efficiency from synthetic observations ($\langle f_{\rm align}\rangle$) to the results from analytical formula ($\langle f_{\rm align}^{\rm ana}\rangle$). Data points show the results from numerical calculations, and solid lines show the running mean values. The dashed line shows the one-to-one correlation. A good correlation between $\langle f_{\rm align}\rangle$ and $\langle f_{\rm align}^{\rm ana}\rangle$ despite different models of grain alignment.}
    \label{fig:falign}
\end{figure}

To compare with synthetic results, we also calculate the mass-weighted alignment degree directly through Equation (\ref{eq:falign_mass}), denoted by $\langle f_{\rm align}^{\rm ana}\rangle$, using the values of $a_{\rm align}$ and $\delta_{\rm mag}$ from Equations (\ref{eq:aalign_ana}) and (\ref{eq:delta_m}), respectively, with the input parameters of the mean gas density $\langle n_{\rm H}\rangle$ and $B$ from MHD simulations. Figure \ref{fig:falign} shows the comparison of $\langle f_{\rm align}\rangle$ obtained from synthetic observations and $\langle f_{\rm align}^{\rm ana}\rangle$ obtained from our analytical calculations for the different models of grain alignment. Regardless of grain magnetic properties, the results show a good agreement between synthetic alignment from POLARIS and analytical calculations. However, as seen in Figure \ref{fig:falign}, $\langle f_{\rm align}\rangle$ is slightly higher than $\langle f_{\rm align}^{\rm ana}\rangle$. This is because, in the synthetic modeling by POLARIS, we take the variation of the anisotropic degree of interstellar radiation fields from 0.8 in the outer regions to 0.2 in the inner regions of the cloud rather than the standard ISM anisotropic degree of 0.1 in the entire cloud as we adopt in analytical calculations. Therefore, the analytical formula can be used to calculate the mass-weighted alignment efficiency of grains. 

\subsection{Synthetic Polarization Versus Analytical Polarization Model}
\begin{figure*}
    \centering
    \includegraphics[width = 1\textwidth]{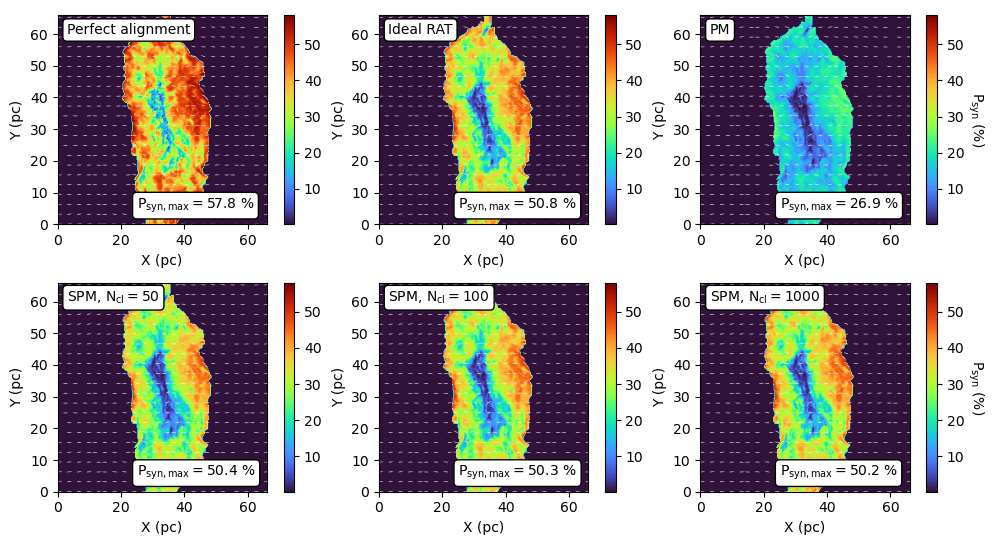}
    \caption{Map of the synthetic polarization degree from POLARIS for the different grain alignment models and the mean B-field in the POS (viewing angle $\gamma_{\rm view}=90^{\circ}$). The polarization degree from the RAT and MRAT mechanisms is maximum in the outer region and significantly decreases toward the central region of high gas density due to the loss of alignment efficiency by increasing gas collision. The polarization degree is lowest for PM grains and increases with the level of iron inclusions embedded in SPM grains.}
    \label{fig:p_obs}
\end{figure*}

Figure \ref{fig:p_obs} shows the maps of synthetic dust polarization at the wavelength of $870\,\rm\mu m$ obtained from the POLARIS code for different grain alignment models, assuming the mean B-field in the POS ($\gamma_{\rm view}=90^{\circ}$). For the perfect alignment model, the polarization degree is highest and can reach a maximum value of $P_{\rm syn, max} \approx 57.8\%$ at the outer region. However, it decreases considerably in the central region (blue area) due to the depolarization effect caused by magnetic turbulence. For the model of Ideal RAT, the polarization degree is lower than the PA model due to the lack of alignment of small grains, which becomes more important toward dense regions. As a result, the polarization degree decreases from the maximum value of $50.8\%$ in the outer and diffuse region to $\sim 5\%$ at the central region of highest density. 

For the model of PM grains, the overall polarization degree reduces with $p_{\rm syn, max} \approx 26.9\%$ due to the lower alignment efficiency. The polarization degree is significantly enhanced for grains with high iron levels and can recover to the Ideal RAT case when $N_{\rm cl}$ reaches $10^3$ with $P_{\rm syn, max} \approx 50.2\%$. In particular, compared to the PA model, the alignment models based on RATs predict a more prominent polarization hole in the center of the filament (blue areas), which is caused by the loss of RAT alignment (i.e., increase in $a_{\rm align}$) due to the increase of gas randomization and the strong attenuation of the ISRF.


\begin{figure*}
    \centering
    \includegraphics[width = 1\textwidth]{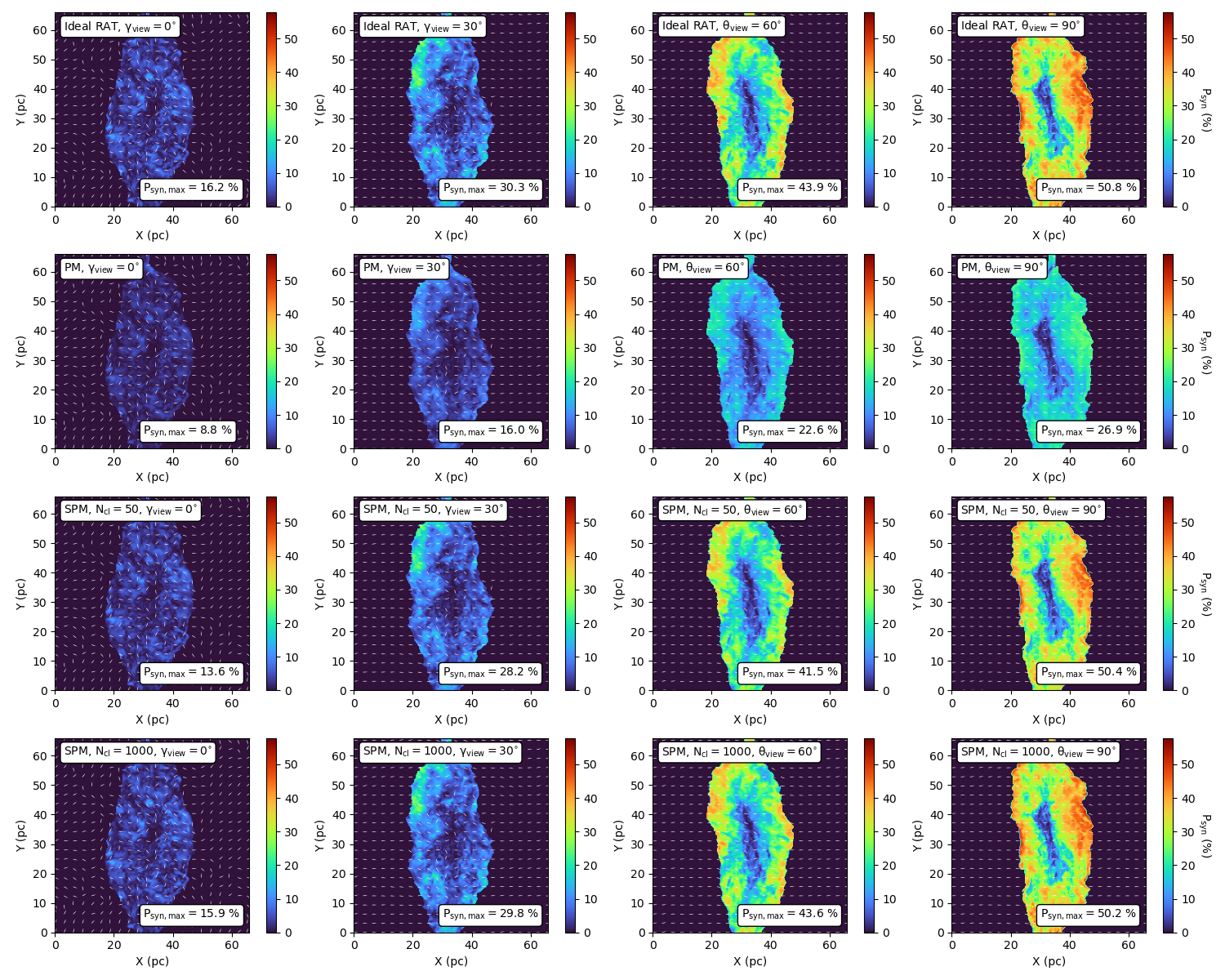}
    \caption{Same as Figure \ref{fig:p_obs}, but for the different viewing angles between the mean B-field and the LOS, $\gamma_{\rm view}$. The polarization degree decreases with decreasing the inclination angle $\gamma_{\rm view}$ due to the projection effect.}
    \label{fig:p_obs_incl}
\end{figure*}

Figure \ref{fig:p_obs_incl} shows the resulting maps of polarization degree, but for different viewing angles, $\gamma_{\rm view} = 0^{\circ}$ (i.e., the mean B-field parallel to the LOS), $30^{\circ}, 60^{\circ}$ and $\gamma_{\rm view} = 90^{\circ}$ (i.e., the mean B-field perpendicular to the LOS). One can see that the polarization degree is significantly reduced from the optimal values at $\gamma_{\rm view}=90^{\circ}$ to the smallest values at $\gamma_{\rm view}=0^{\circ}$ due to the projection effect. Note that the polarization degree is still considerable of $P_{\rm syn, max} < 16\%$ when the mean B-field is along the LOS ($\gamma_{\rm view}=0^{\circ}$) because the local B-fields are not along the LOS due to magnetic turbulence.



\begin{figure*}
    \centering
    \includegraphics[width = 0.48\textwidth]{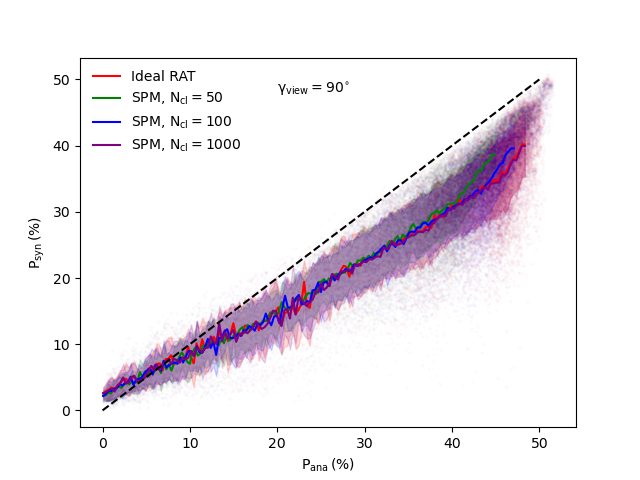}
    \includegraphics[width = 0.48\textwidth]{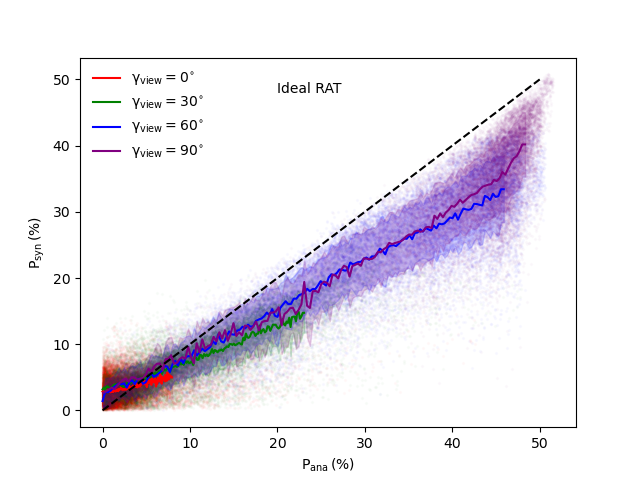}
    \caption{Left panel: Polarization degree from synthetic observations ($P_{\rm syn}$) vs. numerical calculations with the analytical formula, $P_{\rm ana}$ given by Equation (\ref{eq:pol}), for the different grain alignment models, assuming the viewing angle $\gamma_{\rm view} = 90^{\circ}$. Data points show the results from numerical calculations, and solid lines show the running mean values. Right panel: For different viewing angles and Ideal RAT alignment. The dashed line shows the power law of the slope of $1$ when $P_{\rm syn} = P_{\rm ana}$. Both results show a good correlation, but $\rm P_{\rm syn}$ tends to be lower than $\rm P_{\rm ana}$.}
    \label{fig:pnum_pmod}
\end{figure*}

To test whether our physical polarization model can adequately describe the synthetic polarization, we calculate the polarization degree, $\rm P_{\rm ana}$, from Equation (\ref{eq:pol}), by using $p_{i}$ (determined by the maximum polarization for the PA model with $\gamma_{\rm view}=90^{\circ}$, i.e., optimal conditions for polarization), the mean alignment efficiency $\langle f_{\rm align}\rangle$ obtained from our numerical simulations and the mean inclination angle of B-fields $\langle \gamma_{\rm B}\rangle$ from MHD simulations.

The left panel of Figure \ref{fig:pnum_pmod} compares the synthetic polarization degree obtained from POLARIS with the polarization degree calculated by the analytical formula given by Equation (\ref{eq:pol}) for various models of grain alignment and $\gamma_{\rm view}=90^{\circ}$. The right panel shows the same results but for different viewing angles of $\gamma_{\rm view} = 0^{\circ}, 30^{\circ}, 60^{\circ}$ and $90^{\circ}$ and the model of Ideal RAT alignment. In all cases, a good correlation between $\rm P_{\rm syn}$ and $\rm P_{\rm ana}$ is observed, implying that the analytical formula could be used to describe the synthetic polarization. However, the results obtained from synthetic observations tend to be lower than those from analytical calculations. This is caused by the fact that the synthetic polarization can account for the effect of B-field fluctuations relative to the mean B-field along the LOS, whereas the analytical model assumes a uniform B-field along the LOS. This issue will be studied in Section \ref{sec:turbulence}.


\subsection{Inferring inclination B-field angles from synthetic polarization}
As shown in Figure \ref{fig:pnum_pmod}, the analytical dust polarization model (Eq. \ref{eq:pol}) can describe the synthetic polarization. We now apply our method in Section \ref{sec:method} to infer the inclination angles and compare them with the actual angles from MHD simulations.

\begin{figure*}
    \centering
    \includegraphics[width = 1\textwidth]{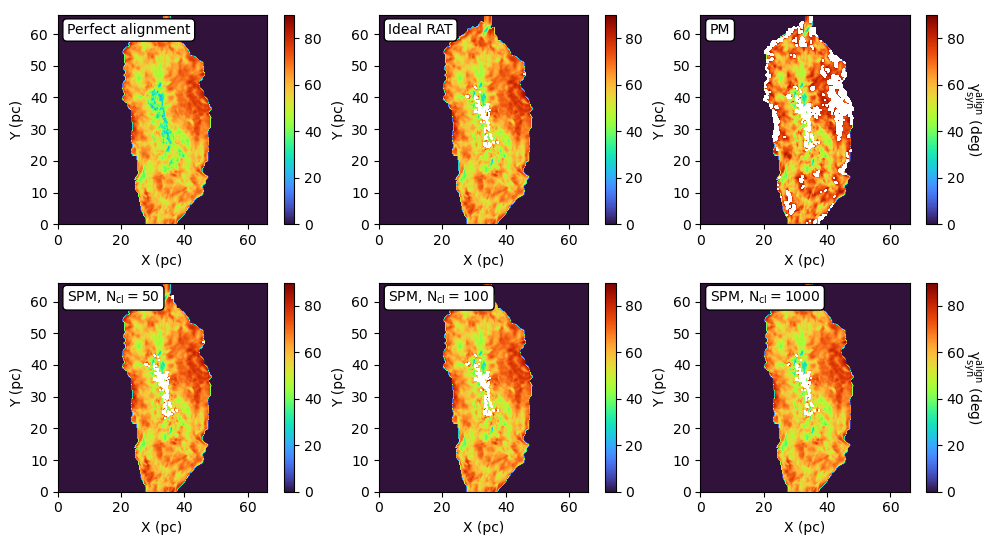}
    \caption{Maps of the inclination angle inferred from our technique for different grain alignment models and the viewing angle $\gamma^{\rm view}=90^{\circ}$. The inclination angle is around 60-80 degrees and does not change significantly for the different grain alignment models. Nevertheless, the inclination angles cannot be retrieved toward denser regions due to the significant reduction of grain alignment.}
    \label{fig:inclination}
\end{figure*}

Figure \ref{fig:inclination} shows the map of inclination angles obtained from our synthetic observations ($\gamma^{\rm align}_{\rm syn}$) for the different models of grain alignment, assuming the viewing angle $\gamma^{\rm view}=90^{\circ}$. Generally, the inclination angle is about 60-80 degrees and tends to decrease to 40-50 degrees toward the dense regions due to the effects of magnetic fluctuations on dust polarization degree (see Figure \ref{fig:pnum_pmod}). When applying our method with the variation of grain alignment, the inclination angle does not change considerably. However, it will return NAN values in the densest regions, which is a result of the reduced grain alignment efficiency by enhanced gas randomization and attenuation of ISRF (as predicted in Table \ref{tab:angle_align}). The situation is worse for PM grains with small $\langle f_{\rm align} \rangle < 0.5$, and the value of $\gamma^{\rm align}_{\rm syn}$ cannot be extracted in some regions of the cloud. This can be resolved when grains have iron inclusions with due to their efficient alignment by MRAT, and the inclination angles can be retrieved from the outer to the inner regions of the cloud.

\begin{figure*}
    \centering
    \includegraphics[width = 1\textwidth]{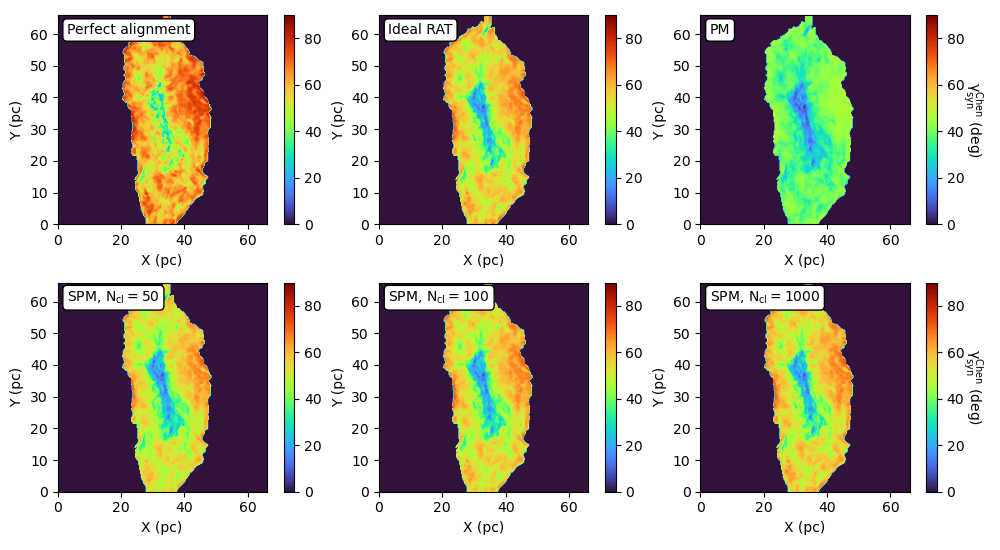}
    \caption{Same as Figure \ref{fig:inclination_Chen}, but being retrieved from Chen's technique with the assumption of uniform grain alignment. The local inclination angles are lower than those from our method and increase with increasing levels of iron inclusions inside SPM grains.}
    \label{fig:inclination_Chen}
\end{figure*}

 Using the synthetic polarization data, we also apply the Chen technique and infer the mean inclination angle for each grain alignment model, denoted by $\gamma_{\rm syn}^{\rm Chen}$, which is shown in Figure \ref{fig:inclination_Chen}. In comparison with our method, the inclination angles from Chen's method vary significantly with the magnetic properties of grains, with the increased inclination angle for SPM grains with high levels of embedded iron. 

To quantify the quality of our method, we compare the inclination angles of B-fields extracted from synthetic dust polarization using both our method and Chen's method with the true inclination angles from MHD simulations. For each cell of the simulation box, the true inclination angles of B-fields are defined as
\bea
\sin{\gamma_{\rm B_{(i,j,k)}}}  = \sqrt{\frac{B_{x_{(i,j,k)}}^2 + B_{y_{(i,j,k)}}^2}{|B|_{(i,j,k)}^2}},\label{eq:gamma_B}
\ena
and the mean inclination angle is calculated by
\bea
\sin{\langle \gamma_{\rm B}} \rangle = \sqrt{\frac{\overline{B_x^2} + \overline{B_y^2}}{\overline{|B|^2}}}.\label{eq:gamma_meanB}
\ena

The inferred inclination angles can also be compared with the density-weighted of the mean inclination angle along the LOS, denoted by $\langle \gamma_{\rho} \rangle$, which is determined by
\bea
\sin{\gamma_{\rho}}  = \sqrt{\frac{\int_{\rm LOS}\,n_\H \sin^2\gamma_{\rm B_{(i,j,k)}}\,\rm dz}{\int_{\rm LOS}\,n_\H\,\rm dz}},\label{eq:gamma_rho}
\ena
where $\gamma_{\rm B_{(i,j,k)}}$ are the local inclination angles derived from Equation (\ref{eq:gamma_B}).

\begin{figure*}
    \centering
    \includegraphics[width = 0.48\textwidth]{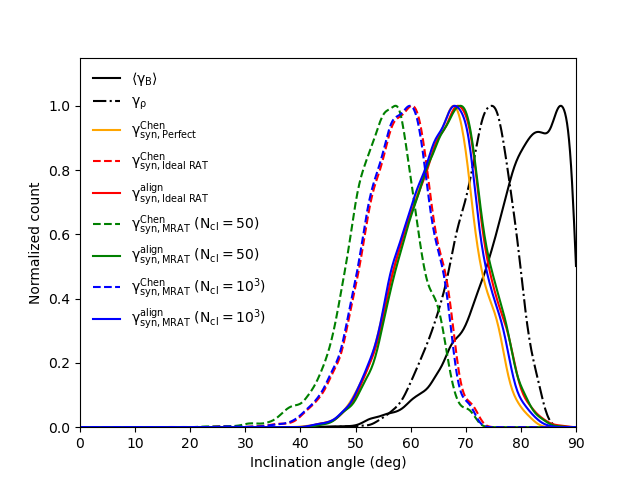}
    \includegraphics[width = 0.48\textwidth]{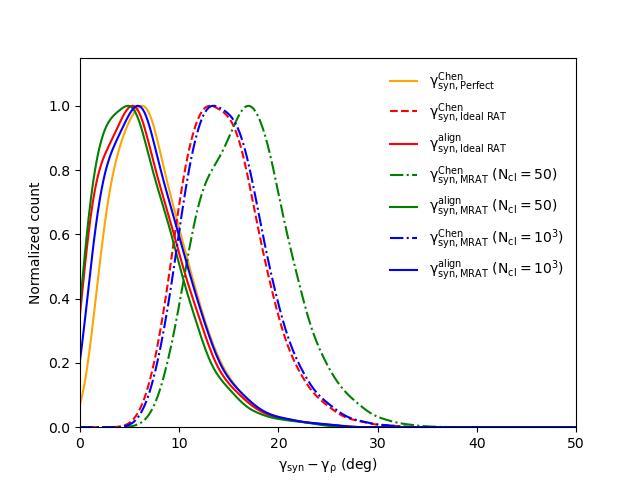}
    \caption{Left panel: Histogram of the inclination angle inferred from our technique and Chen's method for different grain alignment models. Right panel: the offset of the inclination angle to the density-weighted of the mean inclination angle $\gamma_{\rm syn} - \gamma_{\rho}$. The color solid lines show the results from our method, and the color dashed lines for Chen's method. The peak values from our method are less different than the true angle by 5-10 degrees and do not change regardless of grain magnetic properties.}
    \label{fig:histogram}
\end{figure*}

Figure \ref{fig:histogram} shows the distribution of the inclination angles obtained from our method (solid color lines) and Chen's method (dashed color lines), compared with the true angle $\langle \gamma_{\rm B}\rangle$ (solid black line) and the density-weighted angle $\gamma_{\rho}$ (dash-dotted black line). The most probable values of the inclination angles (i.e., the inclination angle at the peak distribution) from each technique are summarized in Table \ref{tab:incl_most}. As one can see, the peak inclination angle from our method tends to be close to the case of perfect alignment (yellow line) and only different by $5-10^{\circ}$, whereas that from Chen's method is lower and differs by $10-20^{\circ}$ from the true value. In particular, the mean inclination angle from our method is insensitive to the grain alignment model and grain magnetic properties. This is expected because we consider the alignment efficiency in deriving the inclination angle.


\begin{table*}
    \centering
    \caption{Summary of the most probable values (i.e., the peak inclination in the histograms) of the inclination angles derived from the true B-fields in MHD simulation, the mean inclination angles from the Chen technique and our method.}
    \begin{tabular}{c c c c c c c c c}
    \toprule
      $\langle \gamma_{\rm B} \rangle^{\bar{\wedge}}$  & $ \gamma_{\rho}^{\bar{\wedge}}$ & $ \gamma_{\rm syn, Perfect}^{\rm Chen^{\bar{\wedge}}}$ & $\gamma_{\rm syn, Ideal\,\,RAT}^{\rm Chen^{\bar{\wedge}}}$ & $\gamma_{\rm syn, MRAT}^{\rm Chen^{\bar{\wedge}}}$ & $\gamma_{\rm syn, MRAT}^{\rm Chen^{\bar{\wedge}}}$ & $\gamma_{\rm syn, Ideal\,\,RAT}^{\rm align^{\bar{\wedge}}}$ & $\gamma_{\rm syn, MRAT}^{\rm align^{\bar{\wedge}}}$ & $\gamma_{\rm syn, MRAT}^{\rm align^{\bar{\wedge}}}$\\

       & & & & $(\rm N_{\rm cl} = 50)$& $(\rm N_{\rm cl} = 1000)$& & $(\rm N_{\rm cl} = 50)$& $(\rm N_{\rm cl} = 1000)$\\
      \midrule
       $87.29^{\circ}$ & $74.67^{\circ}$& $67.81^{\circ}$& $60.24^{\circ}$& $57^{\circ}$& $59.88^{\circ}$& $68.54^{\circ}$& $69.08^{\circ}$& $68^{\circ}$\\
      \bottomrule
    \end{tabular}
    \label{tab:incl_most}
\end{table*}

\begin{figure*}
    \centering
    \includegraphics[width = 1\textwidth]{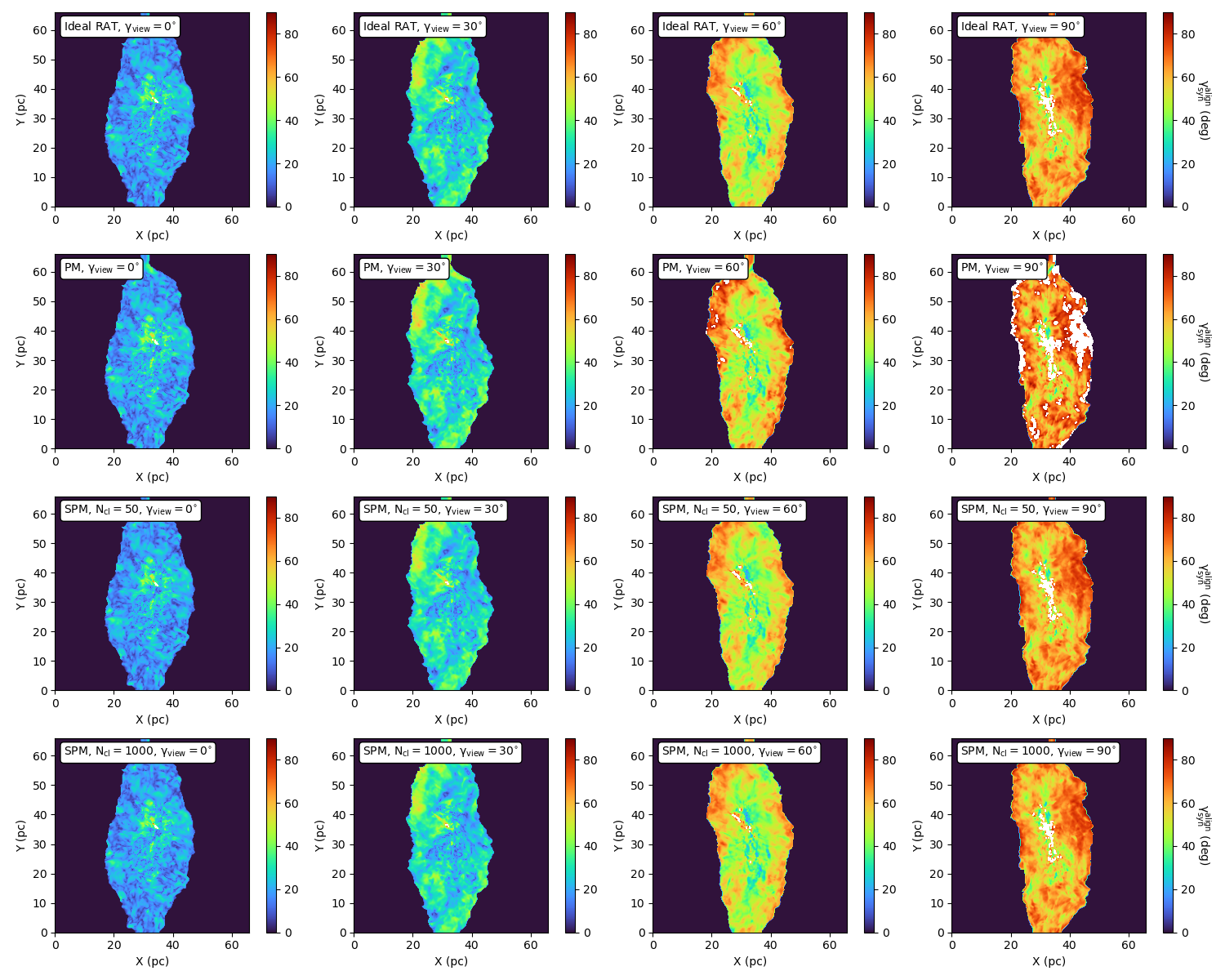}
    \caption{Resulting maps of inclination angles derived from our technique similar to Figure \ref{fig:inclination}, assuming the variation of viewing angles between the mean B-field and the LOS. For smaller $\gamma_{\rm view}$, the significant depolarization is mainly caused by the local inclination angles of B-fields, leading to fewer NAN values.}
    \label{fig:viewing_incl}
\end{figure*}

Figure \ref{fig:viewing_incl} shows the map of inclination angles for the different viewing angles. One can see that, despite the variation in viewing angles, the inclination angles can still be extracted accurately. However, the cloud-scale inclination angles contribute significantly to the depolarization when $\gamma_{\rm view}$ is small (i.e., the B-fields are parallel to the LOS), resulting in fewer NAN values when applying our method in deriving the inclination angles (see also in Table \ref{tab:angle_align}).

\begin{figure*}
    \centering
    \includegraphics[width = 0.48\textwidth]{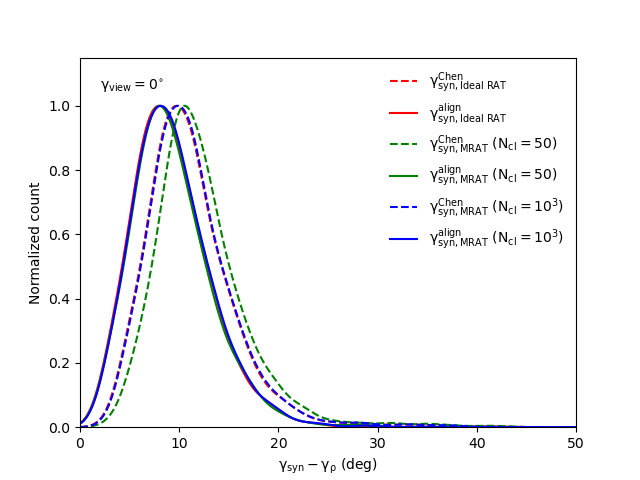}
    \includegraphics[width = 0.48\textwidth]{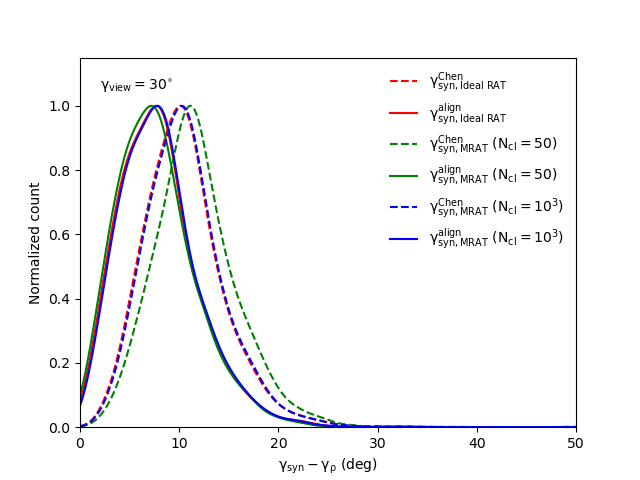}
    \includegraphics[width = 0.48\textwidth]{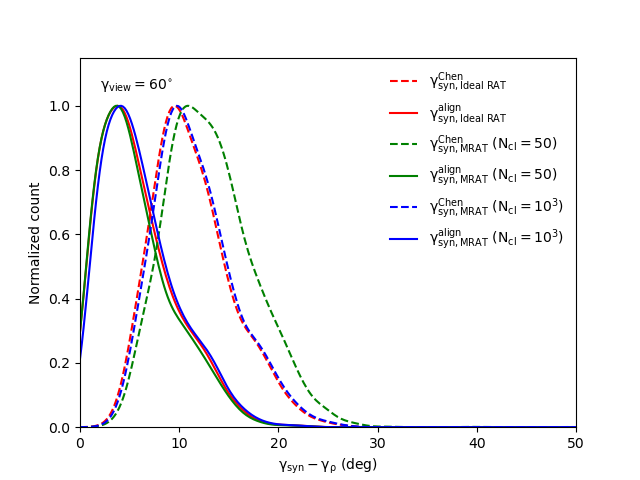}
    \includegraphics[width = 0.48\textwidth]{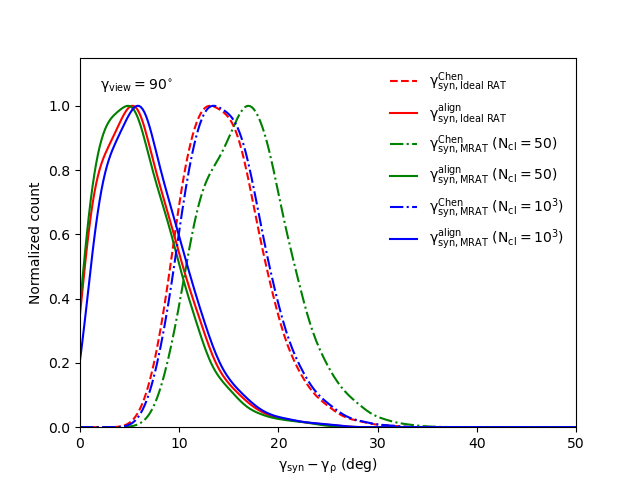}
    \caption{Histogram of the difference of the inferred inclination angles from Chen's and our technique with the actual B-field inclination angles, considering different viewing angles between the mean B-field and the LOS. Even though $\gamma_{\rm view}$ is changing, our method still derives more accurate inclination angles than Chen's technique.}
    \label{fig:Histogram_diff_inclview}
\end{figure*}

Figure \ref{fig:Histogram_diff_inclview} shows the histogram of the offsets of the inclination angles obtained from our method and Chen's method for the different viewing angles (same as the right panel of Figure \ref{fig:histogram}). Table \ref{tab:inclview_most} summarizes all probable values of the inferred inclination angles for the different viewing angles. It is clear that in spite of the change in viewing angles, our method still provides more accurate inclination angles than Chen's method for various models of grain alignment, with a difference of less than $\sim 5-10$ degrees. The inferred inclination angles from Chen's method (dashed lines), on the other hand, strongly depend on grain alignment models and viewing angles, with the difference from the true inclination angle of $10-20$ degrees. The Chen's method becomes less accurate for larger $\gamma_{\rm view}$ as a result of the increasing depolarization effect by grain alignment loss.


\begin{table*}[]
\centering
    \caption{Similar to Table \ref{tab:incl_most} but for different viewing angles between the LOS and the mean B-field}
    \begin{tabular}{c c c c c c c c c}
    \toprule
      $\gamma_{\rm view}$ & $\langle \gamma_{\rm B} \rangle^{\bar{\wedge}}$  & $ \gamma_{\rho}^{\bar{\wedge}}$  & $\gamma_{\rm syn, Ideal\,\,RAT}^{\rm Chen^{\bar{\wedge}}}$ & $\gamma_{\rm syn, MRAT}^{\rm Chen^{\bar{\wedge}}}$ & $\gamma_{\rm obs, MRAT}^{\rm Chen^{\bar{\wedge}}}$ & $\gamma_{\rm syn, Ideal\,\,RAT}^{\rm align^{\bar{\wedge}}}$ & $\gamma_{\rm syn, MRAT}^{\rm align^{\bar{\wedge}}}$ & $\gamma_{\rm syn, MRAT}^{\rm align^{\bar{\wedge}}}$\\

      & & & & $(\rm N_{\rm cl} = 50)$& $(\rm N_{\rm cl} = 1000)$& & $(\rm N_{\rm cl} = 50)$& $(\rm N_{\rm cl} = 1000)$ \\
      \midrule
       $0^{\circ}$ & $11^{\circ}$ & $27.95^{\circ}$& $19.84^{\circ}$& $18.94^{\circ}$& $19.84^{\circ}$& $22^{\circ}$& $22^{\circ}$& $21.82^{\circ}$\\
       
       $30^{\circ}$ & $27.95^{\circ}$ & $40.04^{\circ}$& $30.66^{\circ}$& $28.13^{\circ}$& $30.48^{\circ}$& $33^{\circ}$& $32.46^{\circ}$& $32.82^{\circ}$\\
       
       $60^{\circ}$ & $56.27^{\circ}$ & $62.76^{\circ}$& $49.06^{\circ}$& $47.07^{\circ}$& $48.87^{\circ}$& $54.28^{\circ}$& $55.19^{\circ}$& $54.46^{\circ}$\\
       
       $90^{\circ}$ & $87.29^{\circ}$ & $74.67^{\circ}$& $60.24^{\circ}$& $57^{\circ}$& $59.88^{\circ}$& $68.54^{\circ}$& $69.08^{\circ}$& $68^{\circ}$\\
      \bottomrule
    \end{tabular}
    \label{tab:inclview_most}
\end{table*}

\subsection{Full B-field strength from synthetic polarization}
With our inferred inclination angle, we can calculate the strength of the 3D B-field as follows
\bea
B_{3D}=\frac{B_{\rm POS}}{\sin\gamma},\label{eq:B3D}
\ena
where $B_{\rm POS} = \sqrt{\overline{B_x^2} + \overline{B_y^2}}$ is the B-field strength projected in the plane-of-sky, which is taken from the MHD simulation.

\begin{figure*}
    \centering
    \includegraphics[width = 0.48\textwidth]{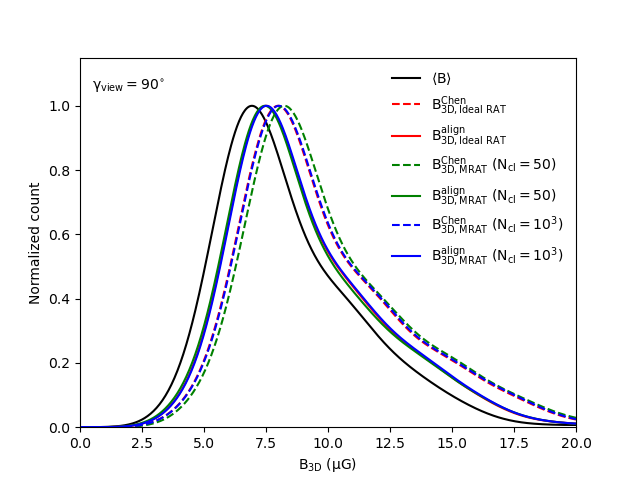}
    \includegraphics[width = 0.48\textwidth]{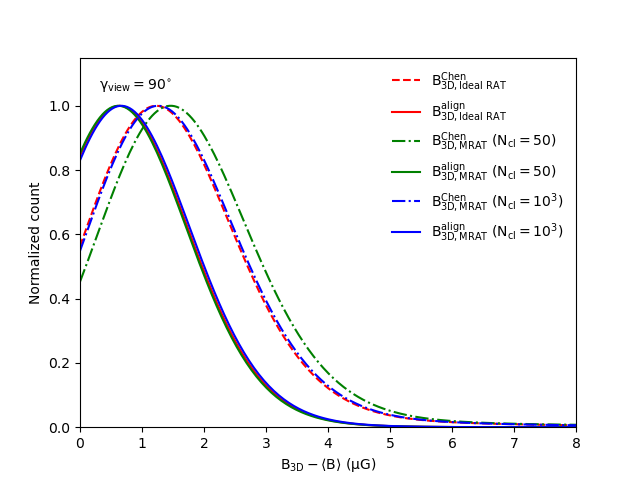}
    \caption{Histogram of the full B-field strength obtained from our method compared to previous methods without uniform grain alignment. The viewing angle of $\gamma_{\rm view}=90^{0}$ is assumed.}
    \label{fig:3DBfield}
\end{figure*}

Figure \ref{fig:3DBfield} compares the full 3D B-field strength from Equation (\ref{eq:B3D}) with the actual strength $\langle B \rangle=7.5\mu G$ from the MHD simulation. One can see that, by taking the inclination angles from our technique, the calculated 3D field strength differs from the true B-field strength by $\rm \sim 0.5-1\,\mu G$, which is only about $10-20\%$ of the actual value, and the inferred value is independent of the grain alignment models. Meanwhile, the lower inclination angles derived from Chen's technique (Figure \ref{fig:histogram}) lead to higher values of 3D B-field strength, with the difference of $\rm \sim 2-3\,\mu G$, i.e., about $40-60\%$ from the true value (see right panel). 

\section{Effects of magnetic turbulence on dust polarization and inferred inclination angles}\label{sec:turbulence}
\subsection{Physical dust polarization model with magnetic turbulence}
In Section \ref{sec:method}, we focus on the effects of grain alignment on the polarization degree by assuming the uniform inclination angle of the mean B-field along the LOS. In reality, the magnetic turbulence induces fluctuations in the local B-field and causes an additional depolarization effect. In the presence of magnetic turbulence, the polarization cross-section in the observer's $\xhat\yhat$ coordinate system (see Figure \ref{fig:JB_Cxy}) can be written as 
\bea
C_{y}-C_{x}&=&C_{\rm pol}f_{\rm align}\sin^{2}\gamma F_{\rm turb},\label{eq:Cpol_turb}\\
\frac{C_{y}+C_{x}}{2}&=&C_{\rm ext}+C_{\rm pol}f_{\rm align}F_{\rm turb}(2/3-\sin^{2}\gamma),\label{eq:Cpol_xy_turb}
\ena
where $\Delta \theta$ is the deviation angle between the local B-field ($\bB$) and the mean field ($\bB_0$), and
\bea
F_{\rm turb}=\frac{1}{2} 
\left[3\langle \cos^{2}(\Delta \theta)\rangle -1\right],\label{eq:Fturb}
\ena
describes the effect of magnetic turbulence on the polarization degree \citep{Lee.1985}. 

The degree of thermal dust polarization (Eq. \ref{eq:pol}) now becomes
\bea
p(\lambda)=\frac{p_{i} \langle f_{\rm align}\rangle\sin^{2}\gamma F_{\rm turb}}{1-p_{i}\langle f_{\rm align}\rangle F_{\rm turb}(\sin^{2}\gamma-2/3)}.\label{eq:pol_turb}
\ena

Setting $p(\lambda)$ to $p_{\rm obs}$, we obtain the inclination angle
\bea
\sin^{2}\gamma = \frac{p_{\rm obs}(1+2p_{i}\langle f_{\rm align}\rangle F_{\rm turb}/3)}{p_{i}\langle f_{\rm align}\rangle F_{\rm turb}(1+p_{\rm obs})}.\label{eq:chi2_align_turb}
\ena

For small fluctuations of $\Delta \theta <1$ (i.e., sub-\alfvenic turbulence), one can make the approximation 
\bea
F_{\rm turb}=1-1.5\langle \sin^{2}(\Delta \theta)\rangle\approx 1-1.5(\delta \theta)^{2},\label{eq:Fturb_approx}
\ena
where $\delta \theta =\langle (\Delta \theta)^{2}\rangle^{1/2}$ denotes the dispersion of the deviation angle. 

Equation (\ref{eq:Fturb_approx}) reveals that the increase in the B-field angle dispersion $\delta \theta$ (stronger magnetic turbulence) reduces the value of $F$, which decreases the polarization degree. Therefore, if the turbulence is isotropic, the B-field angle dispersion $\delta \theta$ corresponds to $\delta \phi$, but their correlation depends on the viewing angles, as shown in Figure \ref{fig:angdis_vs_F} (see Appendix \ref{sec:Fturb_apdx} for details).



\subsection{Comparison to synthetic dust polarization}
For our synthetic observations of MHD simulations, the deviation angle between the local and the mean B-field is calculated in each cell of the simulation box as
\bea
\Delta\theta = \cos^{-1}\frac{(\bB \cdot \bB_0)}{\|\bB\| \|\bB_0\|},\label{eq:delta_theta}
\ena
where $\bB, \bB_{0}$ are the local and mean B-fields from MHD datacube. From that, we integrate the $\cos^2(\Delta\theta)$ over the LOS and calculate the dispersion angle $\delta \theta$, the factor $F_{\rm turb}$ and the analytical polarization degree as shown in Equation (\ref{eq:pol_turb}).

\begin{figure*}
    \centering
    \includegraphics[width = 0.48\textwidth]{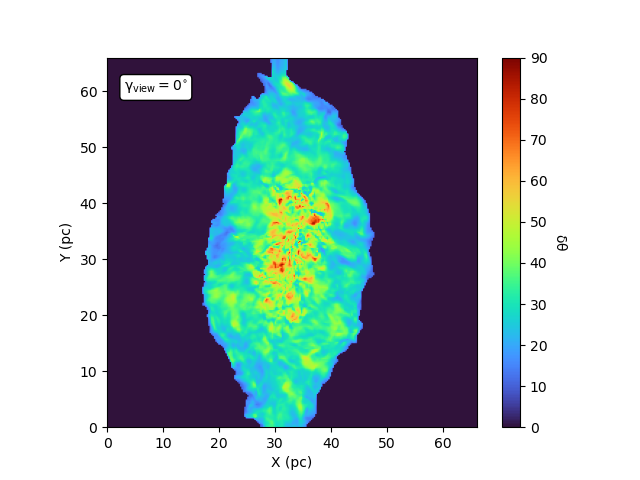}
    \includegraphics[width = 0.48\textwidth]{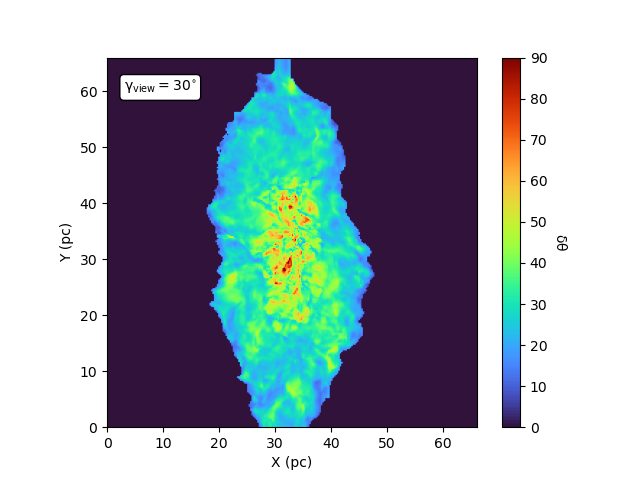} 
    \includegraphics[width = 0.48\textwidth]{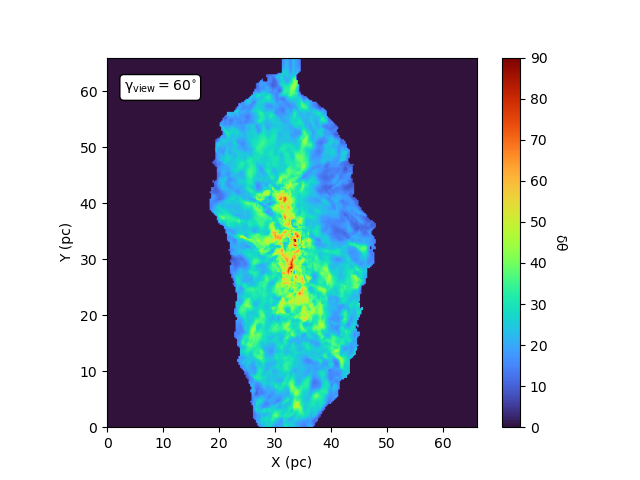}
    \includegraphics[width = 0.48\textwidth]{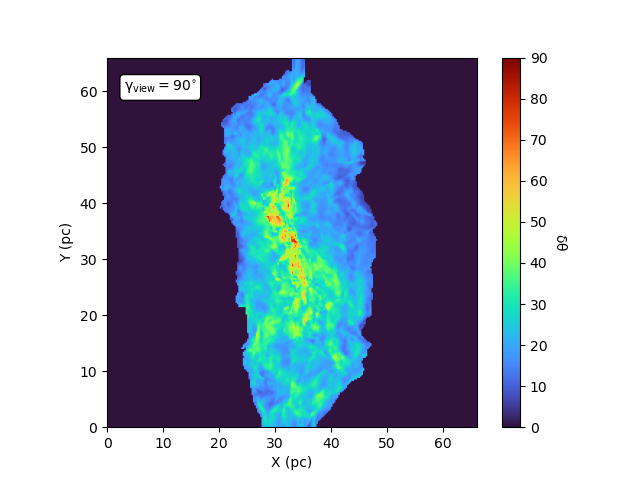}
    \caption{Maps of the fluctuations of the local to the mean B-field, $\delta \theta$, calculated for the different viewing angles. Fluctuations are stronger when the mean B-field becomes closer along the LOS (smaller viewing angles).}
    \label{fig:delta_theta}
\end{figure*}

\begin{figure*}
    \centering
    \includegraphics[width = 0.48\textwidth]{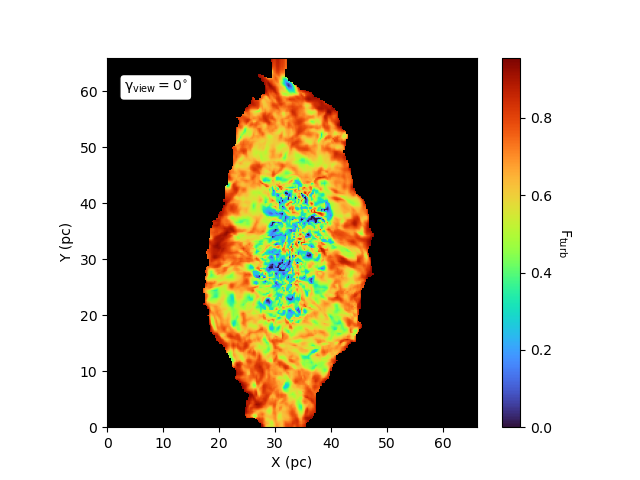}
    \includegraphics[width = 0.48\textwidth]{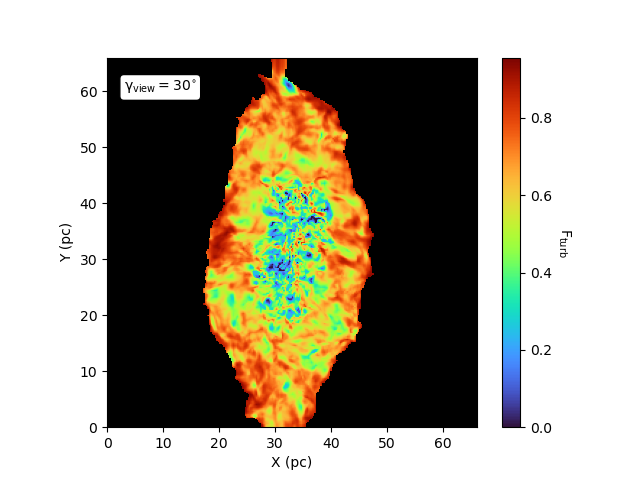}   
    \includegraphics[width = 0.48\textwidth]{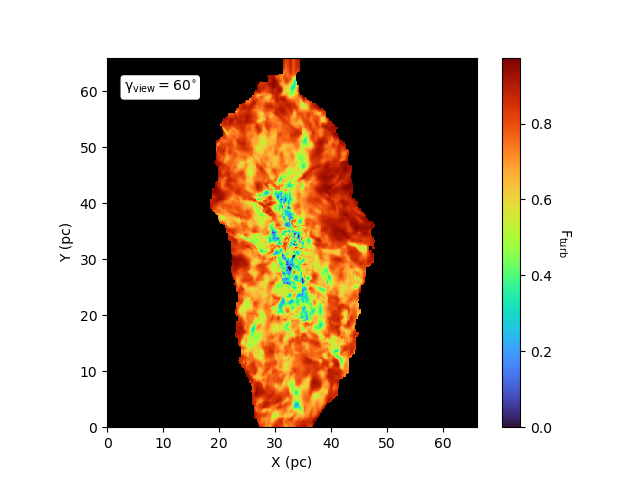}
    \includegraphics[width = 0.48\textwidth]{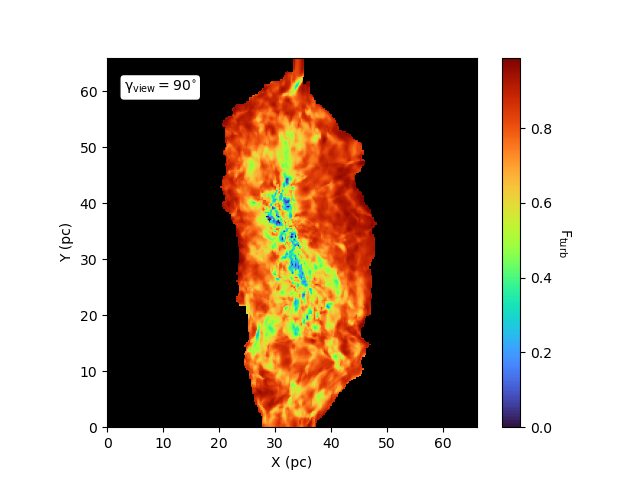}
    \caption{Maps of the term $F_{\rm turb}$ describing the effect of magnetic turbulence on the polarization degree for the different viewing angles. $F_{\rm turb}$ tends to decrease toward the inner region and with the viewing angles due to stronger magnetic fluctuations.}
    \label{fig:F_turb}
\end{figure*}

Figure \ref{fig:delta_theta} shows the resulting maps of the dispersion angle $\delta \theta$ in the entire filamentary cloud, for the different viewing angles. For a given viewing angle, the magnetic fluctuations are weaker in the outer region and tend to increase toward denser regions with $\delta \theta > 50^{\circ}$. Moreover, the fluctuations tend to increase when the viewing angle decreases.

Figure \ref{fig:F_turb} shows the maps of $F_{\rm turb}$ calculated using the angle dispersion maps in Figure \ref{fig:delta_theta}. The value of $F_{\rm turb}$ decreases from the outer to the inner region due to stronger turbulence. Moreover, the depolarization is stronger when the mean field is closer to the LOS, also due to stronger fluctuations.

\begin{figure*}
    \centering
    \includegraphics[width = 0.48\textwidth]{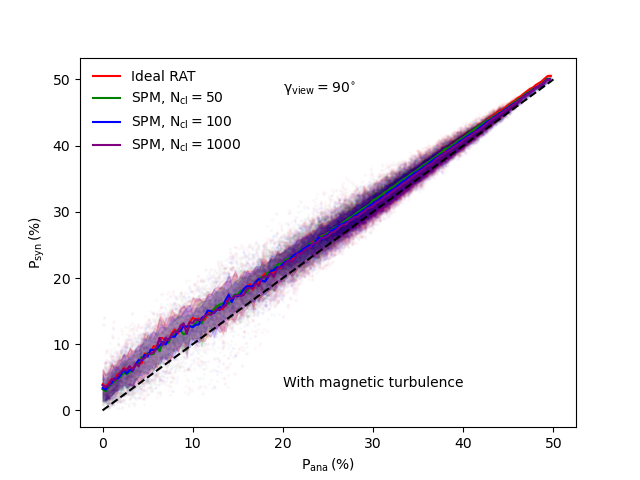}
    \includegraphics[width = 0.48\textwidth]{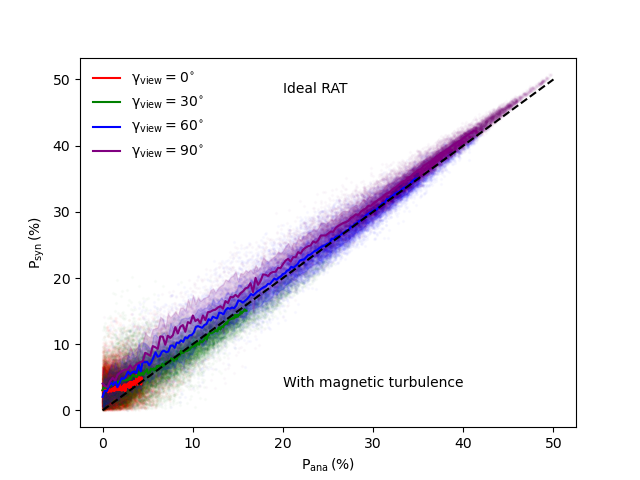}
    \caption{Same as Figure \ref{fig:pnum_pmod} but with the presence of B-field turbulence in the polarization model described by Equation (\ref{eq:pol_turb}). The analytical model becomes in better agreement with synthetic results when the turbulence is included.}
    \label{fig:p_fluc_mag}
\end{figure*}

Figure \ref{fig:p_fluc_mag} (left panel) shows the comparison of the synthetic polarization to the analytical model similar to the previous Figure \ref{fig:pnum_pmod} but in the presence of random B-fields described in Equation (\ref{eq:pol_turb}). In all cases, when the additional term $F_{\rm turb}$ is included, the analytical model becomes in agreement with the synthetic polarization than the model without the turbulence term shown in Figure \ref{fig:pnum_pmod}.  

\subsection{Inferred inclination angles and 3D B-field strengths}
\label{sec:incl_turb}
\begin{figure}
    \centering
    \includegraphics[width = 0.48\textwidth]{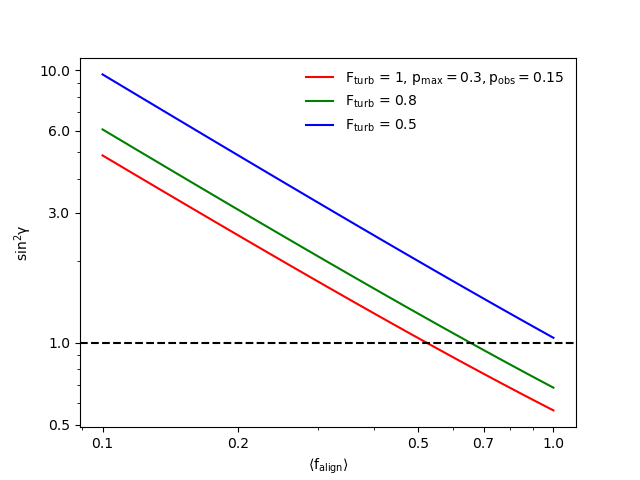}
    \caption{The variation of $\sin^2\gamma$ derived from Equation (\ref{eq:chi2_align_turb}) with respect to the grain alignment efficiency $\langle f_{\rm align} \rangle$ and the turbulence factor $F_{\rm turb} = 1, 0.8, 0.5$, considering $p_{\rm obs} = p_{\rm max}/2 = 0.15$. The inferred inclination angles cannot be extracted when $\langle f_{\rm align} \rangle < 0.5$ or $F_{\rm turb} < 0.5$.}
    \label{fig:falign_incl}
\end{figure}

Following this step, we include the factor $F_{\rm turb}$ for improving the calculation of inclination angles through dust polarization degree, as described in Equation (\ref{eq:chi2_align_turb}). We take the small test demonstrating the dependence of $\sin^2\gamma$ on the grain alignment efficiency $\langle f_{\rm align} \rangle$ and the turbulence factor $F_{\rm turb}$ as shown in Figure \ref{fig:falign_incl}, assuming $p_{\rm obs} = p_{\rm max}/2$. The values of $\sin^2\gamma$ will be larger than 1 when $\langle f_{\rm align} \rangle < 0.5$ (i.e., significant loss of grain alignment) or $F_{\rm turb} < 0.5$ (i.e., high magnetic turbulence), and can return NAN values when extracting the B-field inclination angles.

\begin{figure*}
    \centering
    \includegraphics[width = 1\textwidth]{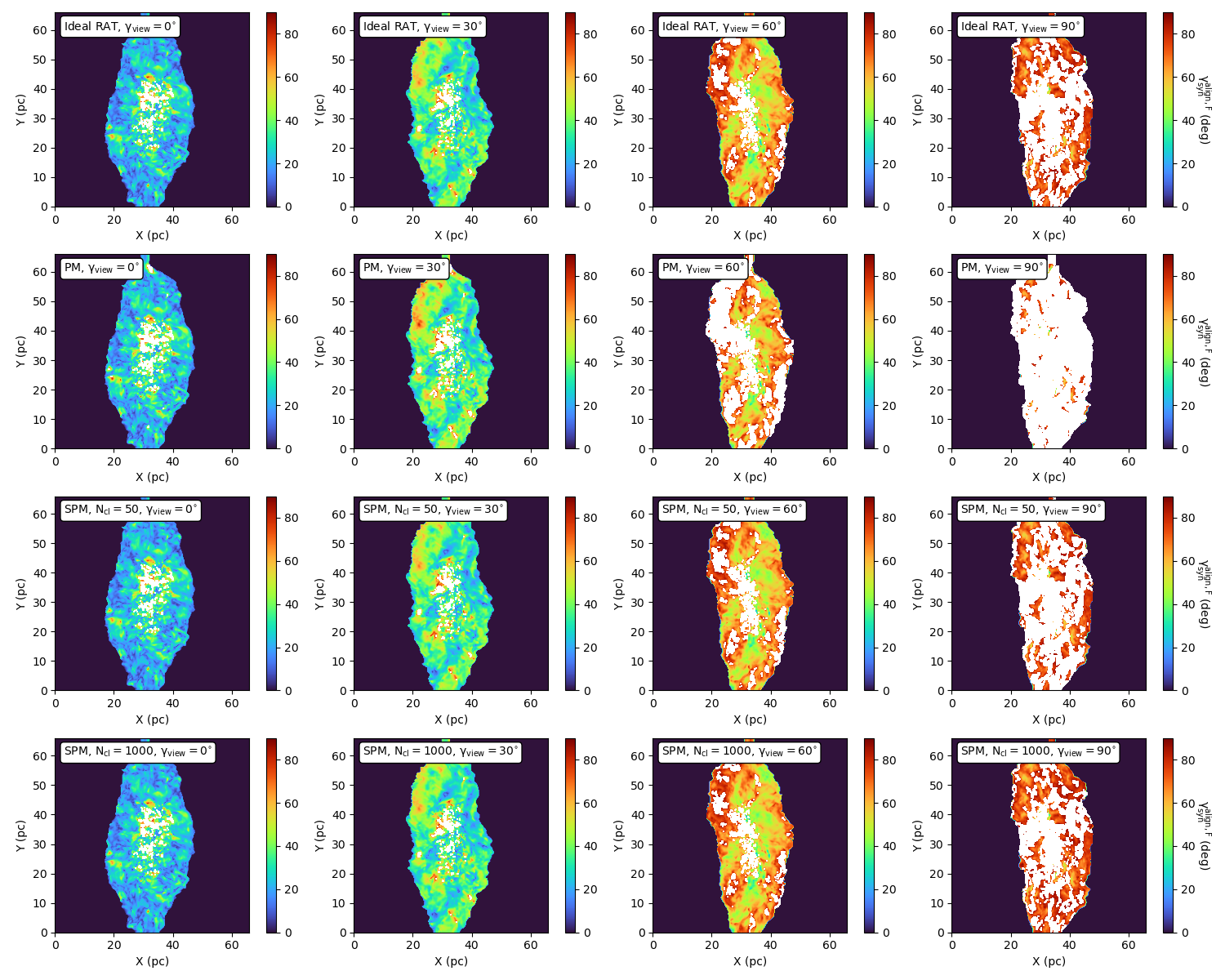}
    \caption{Similar to Figure \ref{fig:viewing_incl} but when the effects of random B-fields along the LOS are considered described by Equation (\ref{eq:chi2_align_turb}). The inferred inclination angles then depend on both random B-fields and the alignment loss, which can result in more NAN values.}
    \label{fig:viewing_incl_F}
\end{figure*}

Figure \ref{fig:viewing_incl_F} shows the results of the inferred inclination angles when $F_{\rm turb}$ is considered, denoted by $\gamma^{\rm align, F}_{\rm syn}$. Compared to Figure \ref{fig:viewing_incl}, the maps show more NAN data points where the thermal dust polarization is dominantly affected by both the magnetic turbulence and the grain alignment loss. As a result, the remaining data points demonstrate the regions in which the magnetic turbulence is weaker, and thus depolarization is mainly by the mean B-field inclination angles and, thus, potentially provide better values of inclination angle. Table \ref{tab:inclview_most_F} summarizes the most probable values of inferred inclination angle, considering the effects of random fields. In comparison to Table \ref{tab:inclview_most}, the inclination angles inferred from our method have improvements and are much closer to the actual values from MHD simulations ($\gamma_{\rho}^{\bar{\wedge}}$). The angles inferred using Chen's method do not change because of their assumption of uniform B-fields.

\begin{table*}[]
\centering
    \caption{Same as Table \ref{tab:inclview_most} but when the effect of magnetic turbulence is taken into account.}
    \begin{tabular}{c c c c c c c c c}
    \toprule
      $\gamma_{\rm view}$ & $\langle \gamma_{\rm B} \rangle^{\bar{\wedge}}$  & $ \gamma_{\rho}^{\bar{\wedge}}$  & $\gamma_{\rm syn, Ideal\,\,RAT}^{\rm Chen^{\bar{\wedge}}}$ & $\gamma_{\rm syn, MRAT}^{\rm Chen^{\bar{\wedge}}}$ & $\gamma_{\rm obs, MRAT}^{\rm Chen^{\bar{\wedge}}}$ & $\gamma_{\rm syn, Ideal\,\,RAT}^{\rm align,F^{\bar{\wedge}}}$ & $\gamma_{\rm syn, MRAT}^{\rm align,F^{\bar{\wedge}}}$ & $\gamma_{\rm syn, MRAT}^{\rm align,F^{\bar{\wedge}}}$\\

      & & & & $(\rm N_{\rm cl} = 50)$& $(\rm N_{\rm cl} = 1000)$& & $(\rm N_{\rm cl} = 50)$& $(\rm N_{\rm cl} = 1000)$ \\
      \midrule
       $0^{\circ}$ & $11.36^{\circ}$ & $29.4^{\circ}$& $19.84^{\circ}$& $18.75^{\circ}$& $19.84^{\circ}$& $25.97^{\circ}$& $23.62^{\circ}$& $26.15^{\circ}$\\
       
       $30^{\circ}$ & $38.4^{\circ}$ & $42.38^{\circ}$& $30.48^{\circ}$& $28.13^{\circ}$& $30.3^{\circ}$& $42.2^{\circ}$& $42.38^{\circ}$& $42.02^{\circ}$\\
       
       $60^{\circ}$ & $67.81^{\circ}$ & $60.6^{\circ}$& $48.7^{\circ}$& $46.71^{\circ}$& $48.33^{\circ}$& $66.91^{\circ}$& $67.09^{\circ}$& $66.19^{\circ}$\\
       
       $90^{\circ}$ & $77.01^{\circ}$ & $73.04^{\circ}$& $56.09^{\circ}$& $51.94^{\circ}$& $55.73^{\circ}$& $79.53^{\circ}$& $79^{\circ}$& $78.81^{\circ}$\\
      \bottomrule
    \end{tabular}
    \label{tab:inclview_most_F}
\end{table*}

\begin{figure*}
    \centering
    \includegraphics[width = 0.48\textwidth]{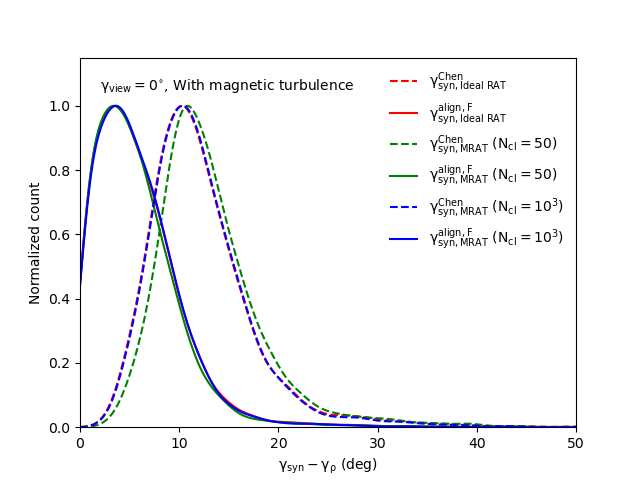}
    \includegraphics[width = 0.48\textwidth]{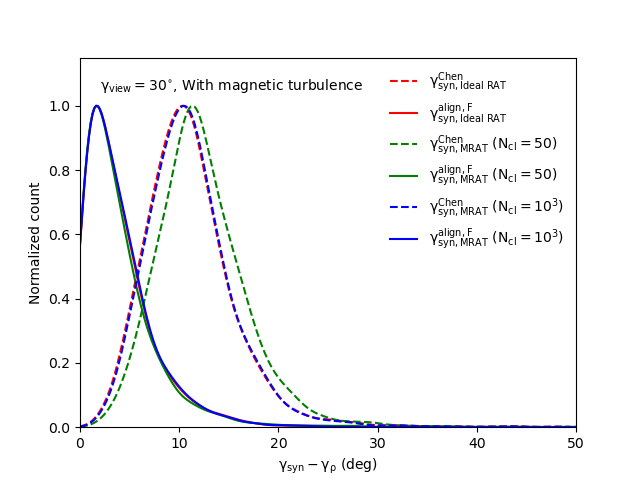}
    \includegraphics[width = 0.48\textwidth]{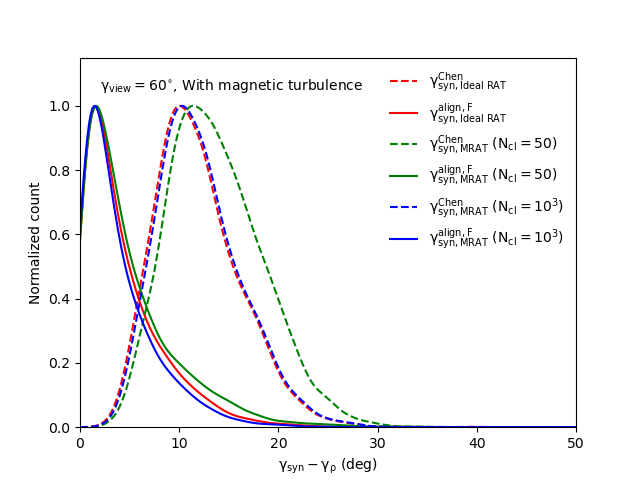}
    \includegraphics[width = 0.48\textwidth]{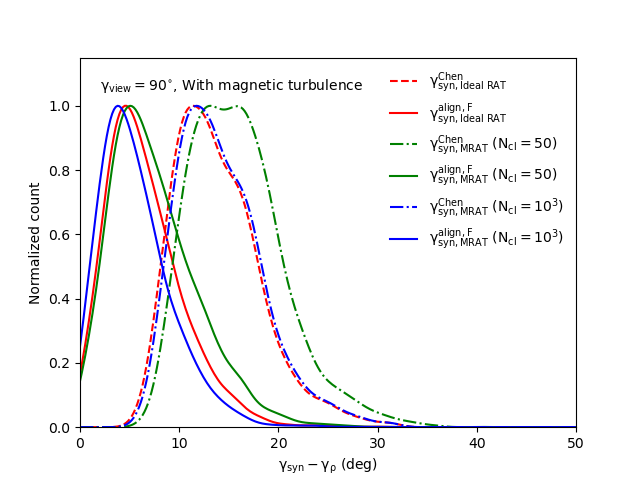}
    \caption{Similar to Figure \ref{fig:Histogram_diff_inclview} but when accounting for the effect of magnetic turbulence. The inferred inclination angles from our method agree better with the true B-field inclination angles from MHD simulations, with the angle difference less than $5^{\circ}$. The results from Chen's method remain unchanged due to their assumption of uniform B-field and alignment.}
    \label{fig:Histogram_diff_inclview_F}
\end{figure*}

Figure \ref{fig:Histogram_diff_inclview_F} shows the difference in the inclination angles inferred from our polarization model when accounting for the effect of turbulence with respect to the actual B-field inclination angles. Compared to Figure \ref{fig:Histogram_diff_inclview}, the inferred inclination angle is much better in agreement with the true inclination angle by only 2-6 degrees in all cases of grain alignment models and various viewing angles.


\begin{figure*}
    \centering
    \includegraphics[width = 0.48\textwidth]{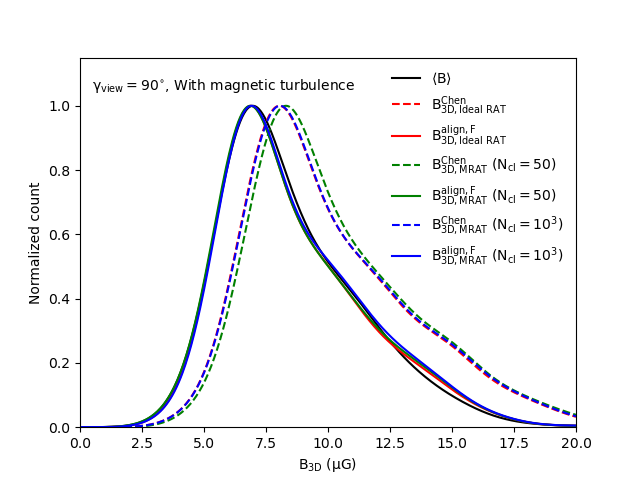}
    \includegraphics[width = 0.48\textwidth]{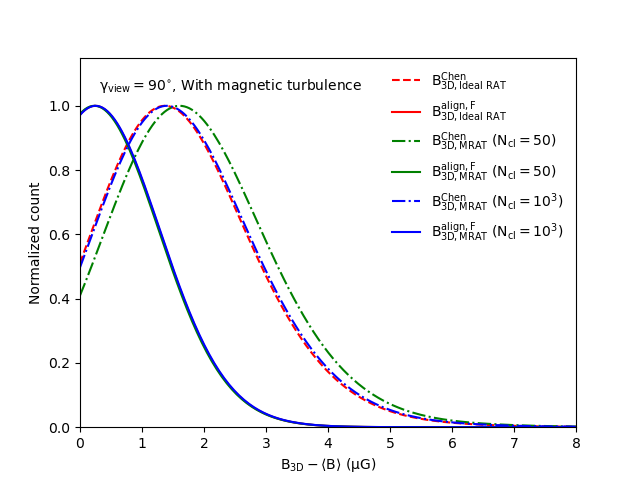}
    \caption{Same as Figure \ref{fig:3DBfield} but when the effect of magnetic turbulence is included in the polarization model and calculations of inclination angles. The 3D B-field strength inferred from our method is much closer to the actual strength of the mean B-field.}
    \label{fig:3DBfield_F}
\end{figure*}

From the improvement of the inferred inclination angles, we calculate again the strength of 3D B-fields, as previously shown in Equation (\ref{eq:B3D}). Figure \ref{fig:3DBfield_F} illustrates the histogram of the 3D B-field strength and its offset when the effect of magnetic turbulence is included. Compared to previous results in Figure \ref{fig:3DBfield}, the derived strength from our method is significantly improved and shows more precisely the mean B-field strength of $\langle B \rangle \sim 5-7.5\,\,\mu\rm G$. The B-field strength inferred from Chen's method remains unchanged.

\section{Discussion}\label{sec:discuss}
\subsection{A physical model of thermal dust polarization based on the MRAT alignment theory}
An accurate, physics-based model of dust polarization is necessary for interpreting polarimetric data and constraining dust physics and B-fields. In this paper, we introduced a physical model of thermal dust polarization, which is the function of four key parameters, including the intrinsic polarization degree ($p_{i}$), mass-weighted alignment efficiency ($\langle f_{\rm align}\rangle$), the mean B-field inclination angle ($\sin^{2}\gamma$), and a term describing the depolarization effect caused by magnetic turbulence, $F_{\rm turb}(\delta \theta)$ (see Eq. \ref{eq:pol} for the model without magnetic turbulence and \ref{eq:pol_turb} for the model with magnetic turbulence). The intrinsic polarization $p_{i}$ is determined by the grain shape elongation. In contrast to previous models that constrain grain alignment using polarization data \citep{Draine.2021no, Hensley.2023} or treat it as an input parameter, here, the value of $\langle f_{\rm align}\rangle$ is calculated from the MRAT alignment theory, which depends on the local gas density, radiation field, grain magnetic susceptibility, and B-field strength \citep{HoangLaz.2016,LazHoang:2021}. The B-field inclination angle describes the projection effect of the mean B-field onto the POS. Finally, the term $F_{\rm turb}$ depends on the nature of magnetic turbulence.

We have tested our physical model of dust polarization degree, both without and with magnetic turbulence, using synthetic polarization observations of MHD simulations for a filamentary cloud with our upgraded POLARIS code (see Section \ref{sec:simul}). We found that the physical polarization model without magnetic turbulence can reproduce the synthetic polarization (see Figure \ref{fig:pnum_pmod}). However, the physical polarization model with magnetic turbulence reproduces better the synthetic polarization than the model without turbulence (see Figure \ref{fig:p_fluc_mag}). Our tested physical polarization model has broad implications for constraining B-fields, dust properties, and grain alignment.



\subsection{Constraining 3D B-fields with dust polarization: comparison of our method to previous studies}
3D B-fields are crucial for understanding the dynamical role of B-fields in the ISM evolution and star formation (\citealt{Hull:2019hw,Pattle.2022}), especially in the formation and evolution of interstellar filaments \citep{HennebelleInutsuka.2019}. However, to date, accurate measurements of 3D B-fields are not yet available. \cite{Chen.20191om} proposed a new method to infer the inclination angle of the B-field using the analytical model of the dust polarization degree. However, the authors adopted an idealized model of dust polarization by assuming that both the B-field and grain alignment efficiency do not change along the LOS. HL23 improved Chen's method by considering the fluctuations of the B-field along the LOS. However, both in Chen and LH23, the grain alignment is assumed to be uniform within MCs, which is unrealistic due to the varying local gas density and radiation fields in MCs.


In this paper, we introduced a general method for constraining the 3D B-field via dust polarization data and using the physical model of dust polarization by incorporating the variation of the grain alignment across the cloud using the modern MRAT alignment theory \citep{HoangLaz.2016,LazHoang:2021}. 
We applied our technique to synthetic polarization data and found that the inferred inclination angles are much closer to the true values of MHD data than Chen's method (see Fig. \ref{fig:Histogram_diff_inclview} and \ref{fig:Histogram_diff_inclview_F}).
Our new technique is particularly useful for tracing 3D B-fields in star-forming regions when the gas density and local radiation field change rapidly. In our follow-up studies, we will apply this method to real observational data of dust polarization to star-forming regions by SOFIA/HAWC+ and JCMT/POL2 and present the results elsewhere.

We also note that both \cite{Chen.20191om} and \cite{HuLaz.2023} derived the polarization $p_{0}$ using the maximum polarization observed from the cloud. This is not well constrained because the value of $p_{0}$ is only determined by the optical conditions for the maximum polarization, $p_{\rm max}$, including the perpendicular B-field and perfect alignment, which is not necessarily satisfied within a cloud. Moreover, the location with observed $p_{\rm max}$ may have an inclined B-field and imperfect alignment. Therefore, there exists the uncertainty in $p_{\rm max}$ and then $p_{0}$. In our present study, the value of $p_{i}$ is directly determined by the grain shape/elongation, which is within a factor 2 different for the grain axial ratio from $1.4-2$ \citep{Draine.2021no}.

In addition, the combination of our method using dust polarization with other B-field measurement techniques can further support probing both B-field and dust properties in star-forming regions. For instance, the Faraday rotation method (see, e.g., \citealt{Tahani.2018}; \citealt{Tahani.2022}) can trace the orientation of B-fields pointing toward or away from us. By combining with our method using dust polarization, we can reconstruct both the morphology and directions of 3D B-fields, particularly in molecular clouds. Also, we note that the inclination angles of the mean B-fields can be potentially inferred by measuring both POS B-fields using dust polarization and LOS B-fields using Zeeman splitting in spectral lines (\citealt{Crutcher:2010p318}) with $\tan\gamma = B_{\rm POS}/B_{\rm LOS}$. By comparing with the inclination angles derived from our technique using dust polarization degree, we can reversely retrieve the information on dust properties such as grain sizes, shapes, and magnetic properties of PM and SPM grains with iron inclusions. This is a new approach to constrain dust properties in star-forming regions, which will be presented in future studies.

\subsection{Can $p\times S$ describe the average grain alignment efficiency?}
The product of the polarization degree ($p$) and the polarization angle dispersion ($S$), i.e., $p\times S$, is usually used to probe the average efficiency of grain alignment using observational polarization data \citep{PlanckCollaboration:2015ev}. Following \cite{Planck.2020}, we calculate the polarization angle dispersion using the second-order structure function of the polarization angles:
\bea
S(\br,\delta)=\left[\frac{1}{N(\delta)}\sum_{i}^{N(\delta)}\left\{\psi(\br+\delta_{i})-\psi(\br)\right\}^{2}\right]^{1/2},\label{eq:Sdef}
\ena
where the summation is taken over $N(\delta)$ pixels within the annulus  centered at $\br$ having the inner radius $\delta/2$ and outer radius $3\delta/2$.

\begin{figure*}
    \centering
    \includegraphics[width = 0.48\textwidth]{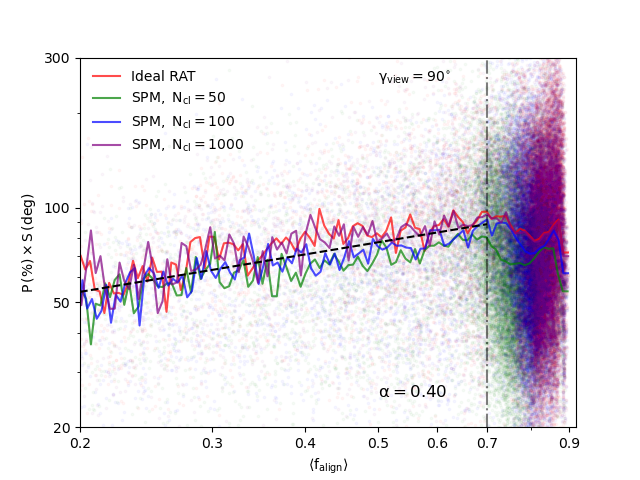}
    \includegraphics[width = 0.48\textwidth]{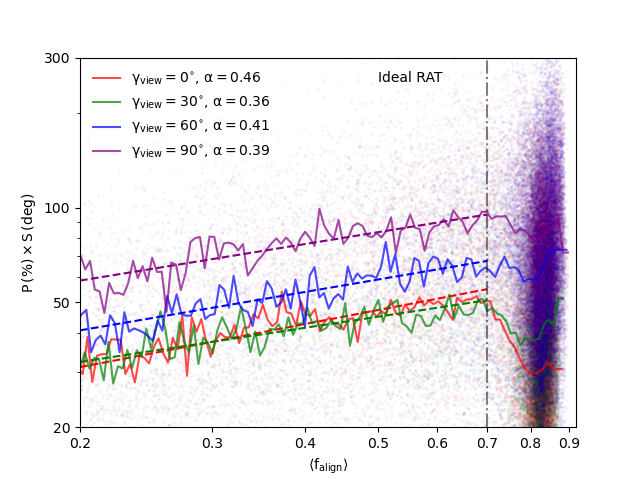}
    \caption{Relationship between $p\times S$ and the mass-weighted alignment efficiency, $\langle f_{\rm align}\rangle$ for the different alignment models (left panel) and viewing angles $\gamma_{\rm view}$ (right panel). The solid lines show the running mean of the product $p\times S$, while the dashed lines show the power-law fit to that running mean. In general, the product $p\times S$ has a good correlation with $\langle f_{\rm align}\rangle$ with a slope of 0.4. However, the product of $p \times S$ strongly fluctuates for higher $f_{\rm align} > 0.7$, which is a result of the reduction of magnetic turbulence toward low-density regions and decreases significantly for lower $\gamma_{\rm view}$.}
    \label{fig:PS_falign}
\end{figure*}

In principle, $S$ describes the level of B-field fluctuations, and larger/smaller $S$ corresponds to stronger/weaker turbulence, which decreases/increases the observed polarization degree. Therefore, the product of $p\times S$ can characterize the average alignment efficiency \citep{PlanckCollaboration:2015ev,Planck.2020}. To test this, we use the polarization degree and mean alignment efficiency from our synthetic observations. We calculate $S$ by choosing the square box of $\delta=3\times 3$ pixels. Figure \ref{fig:PS_falign} shows $p\times S$ vs. $\langle f_{\rm align}\rangle$ for different alignment models (left panel) and various viewing angles (right panel). It shows that $p \times S$ has an overall correlation with the mean alignment efficiency, $\langle f_{\rm align} \rangle$. However, the correlation has a slope of $\alpha\sim 0.4$ for $\langle f_{\rm align} \rangle < 0.7$ in denser regions of the cloud. In lower-density regions with $\langle f_{\rm align} \rangle > 0.7$, the decreased level of magnetic turbulence is more prominent (Figure \ref{fig:delta_theta}) than the increased grain alignment efficiency (Figure \ref{fig:falign_map}), resulting in a downward fluctuation in the product $p \times S$. This value decreases significantly for lower viewing angles $\gamma_{\rm view}$ due to the effect of inclination angles of the mean B-fields. This reveals that the product $p\times S$ may qualitatively describe the alignment efficiency, but it is not recommended to take this product as an accurate indication of grain alignment efficiency. For probing grain alignment with observations, it is suggested to use the mass-weighted alignment efficiency.


\subsection{Dependence of the intrinsic polarization on grain shapes}
The intrinsic polarization $p_{i}=\sigma_{\rm pol}/\sigma_{\rm ext}$ at far-infrared/sub-mm wavelengths much larger than the grain size is determined by the grain elongation \citep{Lee.1985}. For our numerical study, we assumed the grain elongation of $s=2$ that has the intrinsic polarization of $p_{i}\approx 0.58$.
\cite{Draine.2021no} computed the values of $p_{i}$ at $\lambda=850\mum$ for the Astrodust model and found that $p_{i}$ decreases by a factor of $\sim 2$ from $p_{i}\approx 0.58$ at $s=2$ to $p_{i}\approx 0.26$ for $s=1.4$ (see Appendix \ref{sec:astrodust}). Based on their starlight polarization, their constraint for the grain elongation is $s\gtrsim 1.35$.
Moreover, from the RAT theory, we show that the maximum value of $\langle f_{\rm align}\rangle\approx 0.9$ and it cannot reach $1$ because the population of small grains of size $a<a_{\rm align}\sim 0.03-0.05\mum$ are weakly aligned.

We also note that our method of constraining 3D B-fields using dust polarization degree is based on the MRAT alignment theory for composite grains containing embedded iron clusters. The composite grain model is the most likely model of dust in molecular clouds and dense regions of star-forming regions due to the dust evolution via grain-grain collisions. The composite model is also the leading candidate dust in the diffuse ISM \citep{Draine.2021no}. Therefore, our method can be applied to various astrophysical environments, from diffuse to very dense regions. 

Lastly, from observations of dust polarization, one can constrain the intrinsic polarization $p_{i}$ based on the maximum observed polarization degree and the B-field inclination angle and grain alignment efficiency. Nevertheless, due to grain evolution, the grain shape and elongation may change with local environments. However, we expect that the change in grain shapes and then $p_{i}$ is less sensitive than other parameters that affect dust polarization, such as the grain size, alignment degree, and B-fields.

\subsection{Effects of beam size on inferred inclination angles from dust polarization degree}
Above, we have ignored the effect of beam size on the synthetic dust polarization on inferred B-field inclination angles. To examine the effect of beam sizes, we apply a smooth Gaussian beam averaging and convolve with the synthetic polarization data, for three beam sizes of 5, 10, and 15 arcmin. This corresponds to the physical resolutions of 0.15, 0.3, and 0.45 pc, for the case of Orion A located at $\sim 420$ pc (\citealt{Grossschedl2018}). We include the beam convolution effect in the calculations of mean alignment efficiency $\langle f_{\rm align} \rangle$, the depolarization factor by magnetic turbulence $F_{\rm turb}$ and the synthetic polarization degree $P_{\rm syn}$, and recalculate the inferred inclination angles $\gamma_{\rm syn}$ from Equation \ref{eq:chi2_align_turb}.

\begin{figure*}
    \centering
    \includegraphics[width = 1\textwidth]{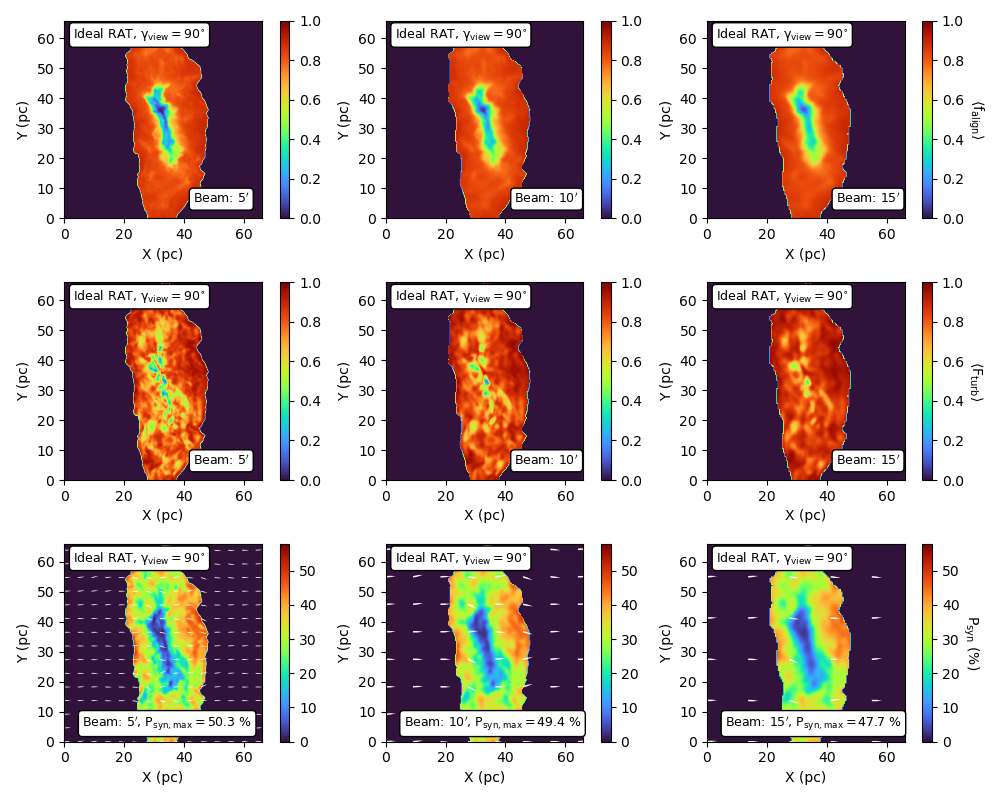}
    \caption{The effects of beam size on the synthetic data of mean alignment efficiency $\langle f_{\rm align} \rangle$ (top), the factor of magnetic turbulence $F_{\rm turb}$ (middle) and the polarization degree $P_{\rm syn}$ (bottom), for the beam size of 5, 10 and 15 arcmin. The Ideal RAT alignment model and $\gamma_{\rm view} = 90^{\circ}$ are assumed. The value of $\langle f_{\rm align} \rangle$ decreases, but $F_{\rm turb}$ increases with the beam size. The synthetic polarization degree, $P_{\rm syn}$, decreases with increasing the beam size due to stronger fluctuations within the beam.}
    \label{fig:P_beam}
\end{figure*}

Figure \ref{fig:P_beam} shows the beam convolution effect on the mean alignment efficiency $\langle f_{\rm align} \rangle$ (top), the factor $F_{\rm turb}$ (middle) and the synthetic polarization degree $P_{\rm syn}$ (bottom), assuming the Ideal RAT alignment model and $\gamma_{\rm view} = 90^{\circ}$. Due to the beam convolution, all features smaller than the beam size are partially averaged out (see, e.g., \citealt{King2018}), which results in the decrease in the mean alignment efficiency $\langle f_{\rm align} \rangle$, but the increase in $F_{\rm turb}$. However, the synthetic polarization degree, $P_{\rm syn}$, decreases with increasing the beam size, which implies that the depolarization caused by the B-field tangling within the beam is stronger for the larger beam.

\begin{figure*}
    \centering
    \includegraphics[width = 0.5\textwidth]{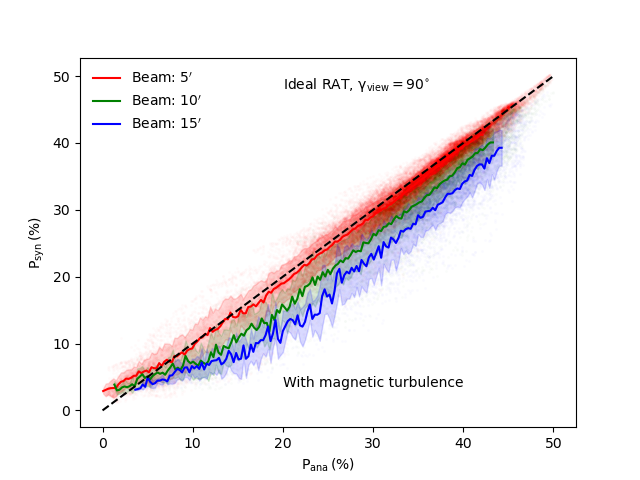}
    \caption{Same as Figure \ref{fig:p_fluc_mag} but when the beam convolution is taken into account, assuming the Ideal RAT alignment model and $\gamma_{\rm view} = 90^{\circ}$. The synthetic polarization $P_{\rm syn}$ decreases with increasing the beam size and is lower than the analytical model $P_{\rm ana}$ due to the fluctuations of B-fields within the beam.}
    \label{fig:p_fluc_beam}
\end{figure*}

Figure \ref{fig:p_fluc_beam} compares the synthetic polarization with the analytical polarization, same as in Figure \ref{fig:p_fluc_mag}, but when the beam convolution effect is taken into account. The Ideal RAT alignment and $\gamma_{\rm view} = 90^{\circ}$ are considered. Due to the depolarization caused by the B-field tangling within the beam, the synthetic polarization degree is essentially lower than the analytical values of $P_{\rm ana}$ obtained by Equation (\ref{eq:pol_turb}) that did not consider the effect of B-field tangling within the beam (see also Figure \ref{fig:pnum_pmod}).

\begin{figure*}
    \centering
    \includegraphics[width = 1\textwidth]{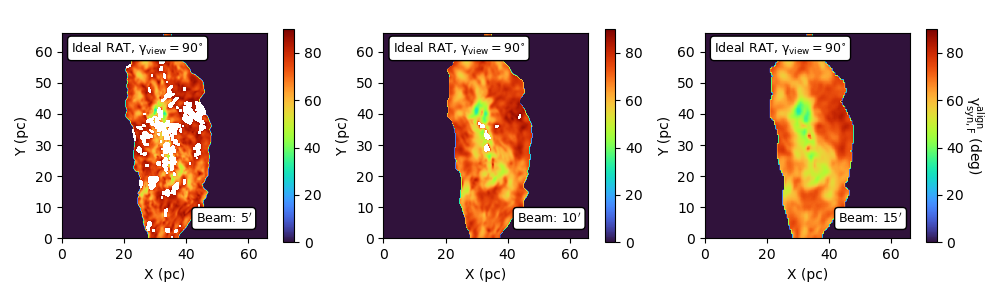}
    \caption{The maps of inferred inclination angles derived from Equation \ref{eq:chi2_align_turb} when the beam convolution effect is considered. Since magnetic turbulence components within the beam are partially destroyed by beam convolution, the effect of magnetic turbulence becomes less significant, which can result in fewer NAN values with increasing beam sizes.}
    \label{fig:incl_beam}
\end{figure*}

\begin{figure*}
    \centering
    \includegraphics[width = 0.48\textwidth]{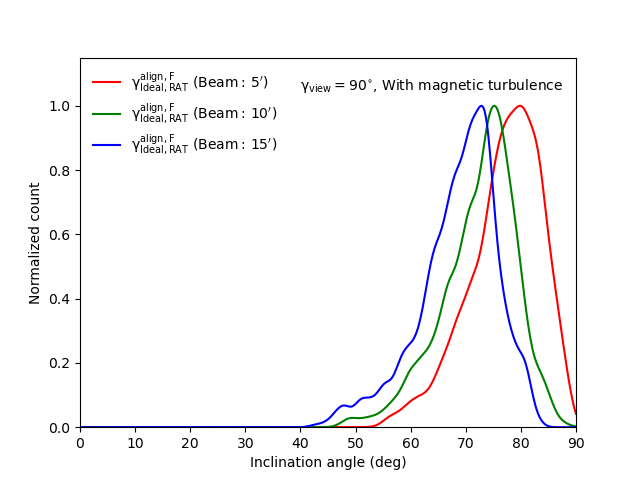}
    \includegraphics[width = 0.48\textwidth]{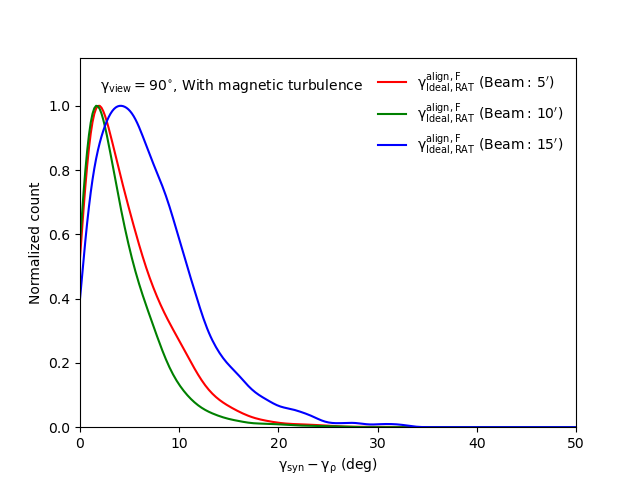}
    \caption{Histogram of the inferred inclination angles in comparison to the true inclination angles from the MHD data, considering the beam convolution effect. The inferred inclination angles become lower and less accurate when observed with larger beams.}
    \label{fig:histogram_beam}
\end{figure*}

Figure \ref{fig:incl_beam} shows the maps of inferred inclination angles $\gamma_{\rm syn}$ when the beam convolution is considered. The distributions of inferred inclination angles for different beam sizes are presented in Figure \ref{fig:histogram_beam}. One can see that for the larger beam size, the inferred inclination angles become lower than the real values and less accurate due to the disregard of the B-field tangling within the beam in the analytical polarization model.\footnote{Our technique could be improved by introducing one more parameter, e.g., $F_{\rm beam}$, that describes additional depolarization caused by B-field tangling within the beam to the analytical model (Eq. \ref{eq:pol_turb}). The value of $F_{\rm beam}$ is smaller for larger beams due to stronger B-field tangling within the beam, which increases the value $\sin^{2}\gamma$ given by Equation (\ref{eq:chi2_align_turb}) and brings the inferred B-field inclination angles closer to the real values. This issue will be quantified in our follow-up studies.}

Note that the choices of arcminute beam are comparable to the Planck beams (\citealt{Planck.2020}), which is applicable in inferring 3D B-fields in molecular clouds on a large scale of $>$ 10 pc. In our follow-up study, we will further examine our technique in tracing 3D B-fields in smaller scales of filaments and dense cores, which is potentially applied in recent single-dish observations from JCMT/POL-2 and SOFIA/HAWC+ with higher spatial resolution.

\section{Summary}\label{sec:summ}
\begin{enumerate}
\item We introduced a physical model of thermal dust polarization using the grain alignment theory based on RATs and magnetic relaxation. At far-IR/sub(-mm) wavelengths, the degree of thermal dust polarization is a function of the intrinsic polarization, the mass-weighted alignment degree, and the B-field inclination.

\item To test our physical polarization model, we performed synthetic polarization observations of MHD simulations for a filamentary cloud using the upgraded POLARIS code with the MRAT alignment theory. We found that the physical polarization model can accurately describe the synthetic dust polarization.

\item From synthetic observations, we found that the mass-weighted alignment efficiency by the MRAT mechanism varies significantly from the outer layer of the filament toward the central region due to the increase in the gas density and decrease of the interstellar radiation field. This reproduces the polarization hole in the synthetic data.

\item Using our tested physical polarization model and synthetic polarization data, we inferred the inclination angle and the strength of 3D B-fields and compared them with the actual values from MHD simulations. Compared to previous methods that disregarded grain alignment, our method with the MRAT alignment provides a more accurate inference of the inclination angle and 3D B-fields.

\item Our physical polarization model with magnetic turbulence can reproduce better the synthetic polarization and enables more accurate inference of the 3D B-fields than the polarization model without magnetic turbulence.

\item Our new method can be applied to real observational data to constrain 3D B-fields using dust polarization data. 

\item When the 3D B-fields are available, our physical polarization model can help constrain the dust properties, such as the grain elongation and magnetic properties, and grain alignment physics.

\end{enumerate}

\acknowledgments
We thank the referee, Yue Hu, for his helpful comments that
improved our paper. We thank Patrick Hennebelle for sharing the MHD datacube with us. This work was partly supported by a grant from the Simons Foundation to IFIRSE, ICISE (916424, N.H.)

\bibliography{ms.bbl}

\appendix
\section{Polarized extinction and polarized emission}\label{sec:apdx}
For the pedagogical purpose, here, we provide detailed formulae for the polarized extinction and emission by aligned grains with the B-field that has an arbitrary orientation in the coordinate system fixed to an observer.

\subsection{Coordinate systems}
Let $\zhat$ be the line of sight, and the plane of the sky is defined by $\xhat\yhat$. To facilitate the calculations of the extinction and polarization cross-section for aligned grains with the B-field, it is convenient to choose the $\xhat$ axis along the B-field projected onto the POS, $B_{\rm POS}$ (see Figure \ref{fig:JB_Cxy}). Calculations of extinction and polarization cross-sections within this coordinate system are previously presented in \cite{Lee.1985} (see also \citealt{PlanckXX.2015}).

\subsection{Polarized extinction}
Unpolarized incident starlight becomes polarized while passing a volume of aligned dust grains. The intensity of transmission starlight with the electric field along $\xhat$ and $\yhat$ is given by, respectively
\bea
I_{x}(\bE\|\xhat)=I_{0}e^{-\tau_{x}},~~I_{y}(\bE\|\yhat)=I_{0}e^{-\tau_{y}},
\ena
where $I_{0}$ is the intensity of unpolarized starlight, and $\tau_{x,y}$ are the optical depth due to starlight extinction by dust ($\tau_{\rm ext}$).

The polarization of transmission starlight is defined by
\bea
p_{\rm ext}=\frac{I_{y}({\bE}\|\yhat)-I_{x}({\bE}\|\xhat)}{I_{y}({\bE}\|\yhat)+I_{x}({\bE}\|\xhat)}=\frac{e^{-\tau_{x}}-e^{-\tau_{y}}}{e^{-\tau_{x}}+e^{-\tau_{y}}}
\ena
For optically thin of $\tau_{x}\sim \tau_{y}\ll 1$, one obtains
\bea
p_{\rm ext}\approx \frac{\tau_{y}-\tau_{x}}{2+\tau_{y}+\tau_{x}}\approx \frac{1}{2}\left(\tau_{y}-\tau_{x}\right),\label{eq:pext1}
\ena
where
\bea
\tau_{x}=N_{\rm H}\int da n_{d}(a)C_{x}(a,\lambda), \tau_{y}=N_{\rm H}\int  n_{d}(a)C_{y}(a,\lambda)da.\label{eq:tauxy}
\ena

Using $C_{x},C_{y}$ from Equations (\ref{eq:Cpol_xy}), one obtains the polarization of starlight by extinction
\bea
\frac{p_{\rm ext}(\lambda)}{N_{\rm H}\sin^{2}\gamma}=\int_{a_{\rm align}}^{a_{\rm max}}da n_{d}(a)C_{\rm pol}(a,\lambda)f_{\rm align}(a).\label{eq:pext2}
\ena

Note that the maximum grain size depends also on the local conditions due to grain growth and destruction processes.

We can write this equation in terms of the averaging:
\bea
\frac{p_{\rm ext}(\lambda)}{N_{\rm H}\sin^{2}\gamma}=\langle f_{\rm align}\rangle \langle \frac{C_{\rm pol}(\lambda)}{V_{\rm gr}}\rangle ,\label{eq:pext}
\ena
where
\bea
\langle f_{\rm align}\rangle &=& \frac{\int_{a_{\rm align}}^{a_{\rm max}}da n_{d}(a)V_{\rm gr}(a)f_{\rm align}(a)}{V_{d}},\\
V_{d}{\langle f_{\rm align}\rangle}\langle \frac{C_{\rm pol}(\lambda)}{V_{\rm gr}}\rangle&=&\int_{a_{\rm align}}^{a_{\rm max}}da n_{d}(a)[C_{\rm pol}(a,\lambda)/V_{\rm gr}(a)]\times V_{\rm gr}(a)f_{\rm align}(a).
\ena

\subsection{Polarized thermal dust emission}
For uniform dust properties within the cloud, the general solution from the radiative transfer equation is 
\bea
I_{\nu}(\tau_{\nu})=I_{\nu}(0)e^{-\tau_{\nu}}+(1-e^{-\tau_{\nu}})B_{\nu}(T_{d}),
\ena
where $\tau$ is the optical depth due to the dust absorption ($\tau_{abs})$ and $d\tau = \rho_{d} \kappa dz$ with $\rho_{d}$ the dust mass density in the cloud, and $\kappa$ the dust opacity.

At long wavelengths of dust thermal emission, the contribution of starlight $I_{\nu}(0)$ is negligible. Therefore, one can write the intensity of thermal dust emission is
\bea
I_{\nu}(\tau_{\nu})\approx (1-e^{-\tau_{\nu}})B_{\nu}(T_{d}).
\ena

For the chosen coordinate system in which $B_{\rm POS}$ along the $\xhat$ axis and grain alignment with the long axis perpendicular to $\bB$ (see Figure \ref{fig:JB_Cxy}), since aligned grains emit radiation with different intensities for $\bE$ along $\xhat$ and $\yhat$, resulting in the polarized thermal emission. The emission is polarized with the polarization vector along $\yhat$. 

The polarization of thermal dust emission is then calculated by
\bea
p_{\rm em}=  \frac{I_{y}-I_{x}}{I_{y}+I_{x}}=\frac{(1-e^{-\tau_{y}})-(1-e^{-\tau_{x}})}{(1-e^{-\tau_{y}})+(1-e^{-\tau_{x}})}.
\ena

For the optically thin case, $\tau_{x}\sim \tau_{y}\ll 1$, one can obtain
\bea
p_{\rm em}=\left(\frac{\tau_{y}-\tau_{x}}{2}\right)\frac{1}{(\tau_{x}+\tau_{y})/2}.\label{eq:pem}
\ena

From Equations (\ref{eq:pext1}) and (\ref{eq:pem}) one can see that $p_{\rm em}\approx p_{\rm ext}/\tau_{abs}$.


\section{Effects of magnetic turbulence on polarization degree}\label{sec:Fturb_apdx}
\subsection{Depolarization factor}
The magnetic turbulence causes the decrease of dust polarization degree (i.e., depolarization), as described by $F_{\rm turb}$, which is given by Equation (\ref{eq:Fturb}). Here we describe the dependence of $F_{\rm turb}$ on magnetic turbulence.

Consider the turbulence is driven by \alfven waves. The random (turbulent) B-field component (${\delta \bB}$) is then perpendicular to the mean field, $B_{0}$, so that the local B-field reads 
\bea
\bB =\bB_{0}+{\delta \bB},\label{eq:Btot}
\ena
which means the tip of $\bB$ is precessing around $\bB_{0}$ with azimuthal angle $\psi=0-2\pi$. Here, we disregard the compressive mode of ${\delta \bB}_{\|}$. 

The dispersion in the deviation angles between the local B-field and the mean field is given by
\bea
\tan(\delta \theta) = \frac{\delta B}{B_{0}}.\label{eq:tan_theta}
\ena

We then have the \alfven Mach number of $M_{\rm A}=\delta v/v_{\rm A}=(\delta B)/B_{0}$. Therefore, we have 
\bea
\tan\delta \theta = M_{\rm A}.
\ena
which yields
\bea
\delta \theta = \tan^{-1} M_{\rm A}.
\ena

Plugging $\delta\theta$ into the depolarization function, we obtain
\bea
F_{\rm turb}(\delta \theta)=\frac{1}{2} \left[3\langle \cos^{2}(\Delta \theta)\rangle -1\right]. \label{eq:Fturb_apdx}
\ena

For strong B-fields and weak turbulence of $M_{\rm A}\lesssim 1$, $\tan\delta \theta\approx \delta \theta=M_{\rm A}$, so
\bea
F_{\rm turb}(\delta \theta)=1-\frac{3M_{\rm A}^{2}}{2}, \label{eq:Fturb_MA}
\ena
which yields the complete depolarization at $M_{\rm A}=(2/3)^{1/2}\approx0.82$.

\subsection{The DCF method and angle dispersion from polarization angles}\label{sec:DCF}
The DCF method \citep{Davis.1951,ChandraFermi.1953} is based on the assumption that the turbulent component of the B-field is equal to the kinetic turbulent energy. Therefore, one can write
\bea
\frac{(\delta B)^{2}}{8\pi }=\frac{\rho \sigma_{v}^{2}}{2},\label{eq:DCF_energy}
\ena
where $\sigma_{v}$ is the dispersion (i.e., standard deviation) of the one-dimensional velocity of the gas.

Using Equation \ref{eq:delta_theta}, $\delta B=\tan(\delta \theta) B_{0}$ \citep{Falceta.2008}, then, one obtains
\bea
B_{0}=\sqrt{4\pi \rho}\frac{\sigma_{v}}{\tan(\delta \theta)}.\label{eq:DCF_B0}
\ena

One can in principle calculate $\delta \theta$ using the dust polarization angle because grains are most likely aligned with the B-field in the ISM and MCs. Let $\phi$ be the polarization angle. The polarization angle dispersion is formally defined by
\bea
\delta \phi\equiv \sigma_{\phi} = \langle (\phi-\bar{\phi})^{2}\rangle,\label{eq:delta_phi_pol}
\ena
where the bracket denotes the averaging over total pixels in a chosen area.

\subsubsection{Angle dispersion from second-order structure function}
In practice, the formula (\ref{eq:delta_phi_pol}) is inaccurate due to the effect of large-scale components of B-fields. Significant efforts are invested to accurately calculate $\delta \phi$ from dust polarization observations. \cite{Hildebrand.2009} suggested the angular dispersion function (ADF) method,
\bea
\langle (\Delta \phi(l))^{2}\rangle^{1/2}=\langle (\phi(\br+\bl)-\phi(\br))^{2}\rangle^{1/2}=\left[\frac{1}{N(l)}\sum_{i=1}^{N(l)} \left\{\phi(\br+\bl)-\phi(\br)\right\}^{2}\right]^{1/2},
\ena
where $l$ is the separation length, and $N(l)$ is the total pixels within the sphere of radius $l$ centered at $\br$ \citep{Falceta.2008}. 

Using the Taylor expansion for the ADF, it yields
\bea
\langle (\Delta \phi(l))^{2}\rangle = b^{2}+m^{2}l^{2}.\label{eq:Delta_phi2}
\ena

The polarization angle dispersion is then
\bea
(\delta \phi)^{2}=\frac{\langle (\delta B)^{2}\rangle}{\langle B^{2}\rangle}\sim \frac{b^{2}}{2-b^{2}},\label{eq:delta_phi2}
\ena
where $(\delta B)=\langle B_{t}^{2}\rangle^{1/2}$ with $B_{t}$ the turbulent component of the B-field \citep{Hildebrand.2009}.

Using polarization angle data, one can calculate the ADF for different lengths $l$ using Equation (\ref{eq:Delta_phi2}). Then, by fitting the resulting data $\langle (\Delta \phi(l))^{2}\rangle$ with Equation (\ref{eq:Delta_phi2}), one can obtain the best-fit parameters $b$ and $m$. Then, using Equation (\ref{eq:delta_phi2}) one derives the angle dispersion $\delta \phi$. By assuming $\delta \theta=\delta \phi$, we can calculate the depolarization parameter $F_{\rm turb}$, which will be used to infer the inclination angle using the polarization degree. The above equation is frequently used to infer the angle dispersion from dust polarization maps (e.g., \citealt{Ngoc.2021,Tram.2023}).

\subsubsection{Angle dispersion from unsharp masking method}
Another method to calculate the map of polarization angle dispersion is using the un-sharp masking method \citep{Pattle.2017}. The angle dispersion is calculated as follows,
\bea
\sigma_{\phi}(\br)=\left[\frac{1}{N(w)}\sum_{i=1}^{N(w)}\left\{\phi_{i}(\br+{\bf w})-\bar{\phi}(\br)\right\}^{2}\right]^{1/2},\label{eq:sigma_phi}
\ena
where $\bar{\phi}$ is the mean angle of the squared box of size $w$ centered at $\br$ and $N(w)$ is the total pixel number within the box.

\begin{figure*}
    \centering
    \includegraphics[width = 0.48\textwidth]{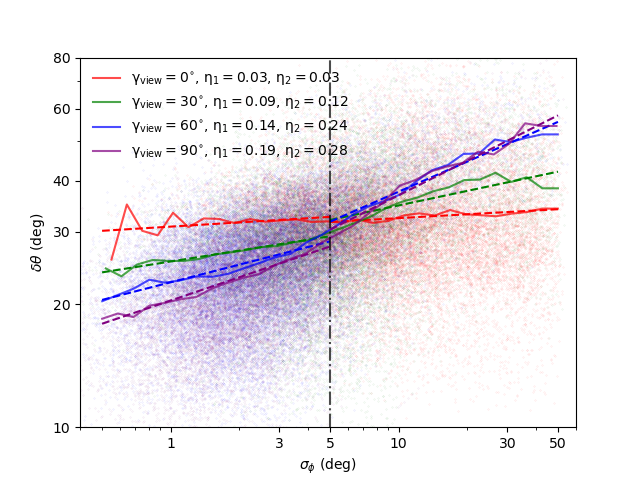}
    \includegraphics[width = 0.48\textwidth]{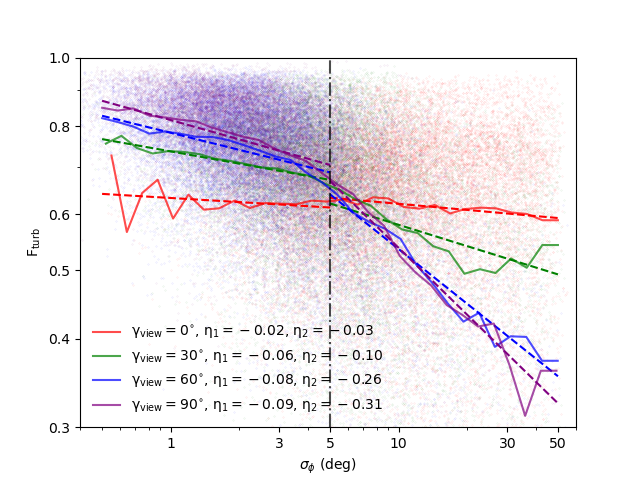}
    \caption{Left panel: Variation of the magnetic fluctuation angle $\delta \theta$ with the polarization angle dispersion $\sigma_\phi$ using the unsharp-masking method for the different viewing angles. Right panel: same as the left one, but for the term $F_{\rm turb}$. Color dashed lines show the results of power-law fit. A correlation between $\delta\theta$ and $\sigma_{\phi}$ or between $F_{\rm turb}$ and $\sigma_{\phi}$ is observed for all cases, except when the mean field is along the LOS.}
    \label{fig:angdis_vs_F}
\end{figure*}

Figure \ref{fig:angdis_vs_F} shows variation of the magnetic angle dispersion $\delta\theta$ (left panel) and $F_{\rm turb}$ (right panel) with the polarization angle dispersion $\sigma_{\phi}$ calculated from the un-sharp masking method for different viewing angles, assuming the squared box size of $3 \times 3$ pixels. The power-law fit is performed to the running means of $\delta\theta$ and $F_{\rm turb}$. Overall, the slope tends to be steeper for larger viewing angles. For small $\sigma_{\phi}$, the slope is rather shallow of $|\eta|<0.1$, which implies the insignificant effect of turbulence on the dust polarization degree. The slope become steeper for $\sigma_{\phi}>5^{\circ}$ with $|\eta|>0.1$, but it varies with viewing angles. For lower viewing angles (i.e., the mean field is along the LOS), the impact of magnetic turbulence is more prominent in the LOS rather than in the POS; consequently, the $F_{\rm turb}$ is less correlated with the dispersion $\sigma_{\phi}$ in the POS with the shallow slope $\eta_{2}\sim -0.03$. The effect of turbulence is more significant in the POS with increasing viewing angles $\gamma_{\rm view} > 30^{\circ}$, which spans $\eta_{2}\sim -0.1$ to $-0.3$. This implies that the factor $F_{\rm turb}$ can potentially be retrieved from the polarization angle dispersion in the POS polarization map, but the effect of the inclination angle itself provides uncertainty.


Note that the angle dispersion is a function of $M_{\rm A}$ with different slopes depending on the mode of MHD turbulence \citep{Lazarian.2022}. In our future papers, we will apply this method to constrain 3D B-fields using dust polarization maps.



\section{Synthetic polarization observations for the Astrodust model and inferred inclination angles}\label{sec:astrodust}
So far, we have emphasized the capabilities of inferring inclination angles from dust polarization using our technique and applied it to the case of the composite model of interstellar dust with the highly elongated oblate shape of $s = 2$. Here, we assume the Astrodust model by \cite{Draine.2021no} with a less elongated oblate shape of $s = 1.4$ and perform synthetic polarization of MHD simulations as in Section \ref{sec:simul}. We apply our technique to the resulting synthetic data to infer the inclination angles.

\begin{figure*}
    \centering
    \includegraphics[width = 1\textwidth]{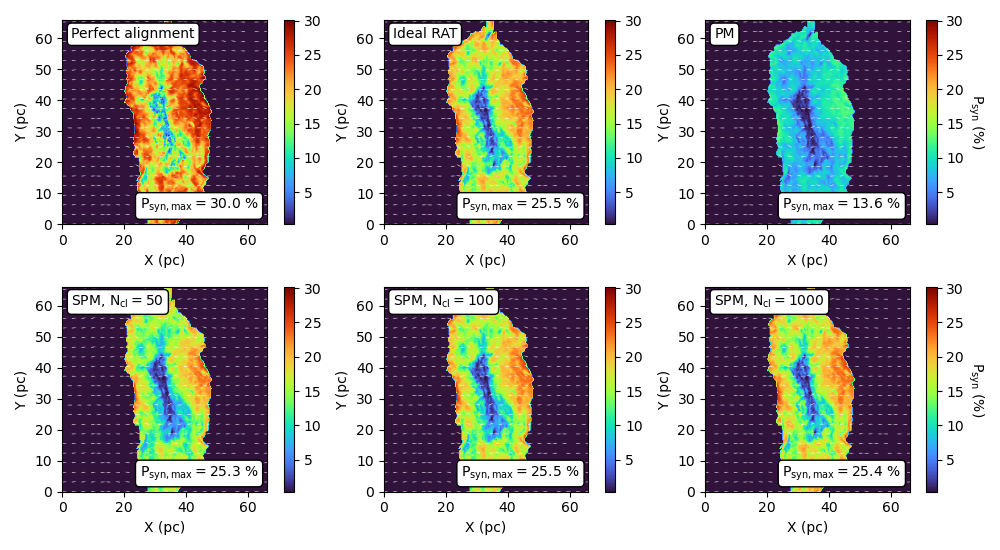}
    \caption{The resulting polarization degree for various grain alignment models similar to Figure \ref{fig:p_obs}, but assuming the Astrodust model with the grain elongation of $s = 1.4$ (\citealt{Draine.2021no}). The viewing angle $\gamma_{\rm view} = 90^{\circ}$ is considered. As grains become less elongated, the overall polarization degree reduces by a factor of 2.}
    \label{fig:Pobs_astrodust}
    \includegraphics[width = 1\textwidth]{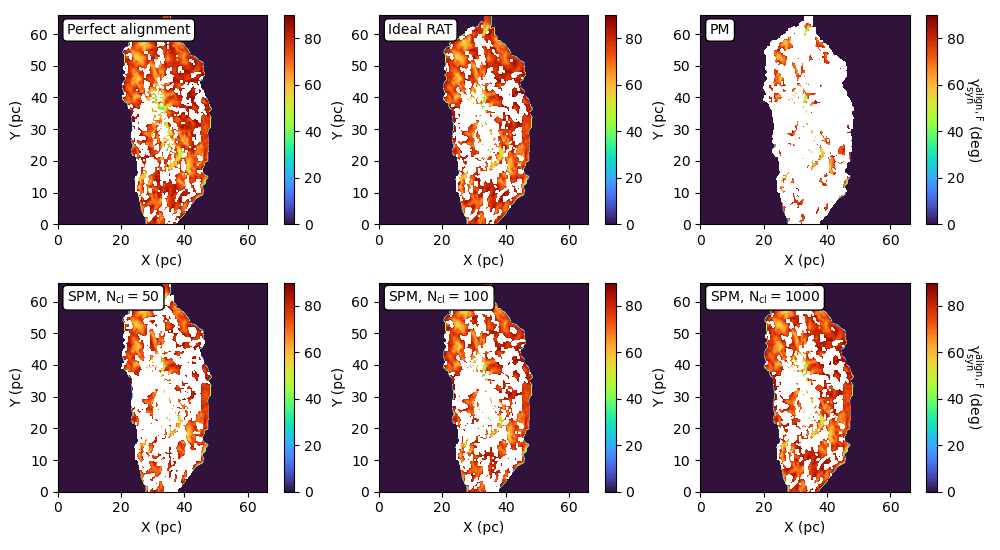}
    \caption{The maps of inferred inclination angles derived from our technique, but when the Astrodust model and less-elongated grains are taken into consideration. The inferred inclination angles do not change in comparison to the previous results in Figure \ref{fig:viewing_incl_F}. }
    \label{fig:incl_astrodust}
\end{figure*}

Figure \ref{fig:Pobs_astrodust} illustrates the synthetic results of polarization degree for the Astrodust model. Since the intrinsic polarization decreases as grains become less elongated, the overall polarization degree for all alignment models reduces significantly compared to the previous results in Figure \ref{fig:p_obs} by a factor of $\sim 2$. For instance, $p_{\rm syn,max} = 50.8\%$ decreases to $p_{\rm syn,max} = 25.5\%$ for the Ideal RAT case - roughly close to the maximum polarization degree found in the diffuse ISM by \cite{Planck.2020}.

Figure \ref{fig:incl_astrodust} shows the maps of the inferred inclinations using our technique when the effects of grain alignment and magnetic turbulence are considered for the Astrodust model and the mean fields perpendicular to the POS (i.e., $\gamma_{\rm view} = 90^{\circ}$). Despite the decreased intrinsic polarization degree due to lower grain elongation, the values of the inferred inclination angle seem to be unchanged compared to the previous calculations illustrated in Figure \ref{fig:viewing_incl_F}. This implies that our technique can be applied to arbitrary grain shapes. 

\section{Application to real polarization data}
In Section \ref{sec:results}, we have applied our new technique to synthetic polarization data to infer 3D B-fields. Here, we outline the main steps to apply it to real polarization data from observations:
\begin{itemize}
\item Assume a dust model of reasonable grain elongation, e.g., constrained by starlight polarization \citep{Draine.2022}, which has an intrinsic polarization $p_{i}$.
\item Derive the map of the gas number density $n_{\rm H}$ from the column density $N_{\rm H}$ and the map of radiation field from dust temperature.
\item Calculate the polarization angle dispersion ($\delta \phi$) and apply the DCF method to calculate the B-field strength, $B_{\rm POS}$ (see Appendix \ref{sec:DCF}).
\item Calculate the minimum size of grain alignment by RATs, the magnetic relaxation using the physical parameters derived from the maps to obtain $f_{\rm align}$ for different grain magnetic properties (e.g., $N_{\rm cl}$). Then, integrate it over the grain size distribution to obtain the map of the mass-weighted alignment efficiency $\langle f_{\rm align}\rangle$.
\item Calculate $F_{\rm turb}$ from Equation (\ref{eq:Fturb_apdx}) using the polarization angle dispersion of $\delta \phi$ (see also Appendix \ref{sec:Fturb_apdx}).
\item With $p_{i}$, $\langle f_{\rm align}\rangle$, $F_{\rm turb}$, and $p_{\rm obs}$ available, the inclination angle is inferred using Equation (\ref{eq:chi2_align_turb}).
\end{itemize}
We will apply this technique to real polarization data from single-dish telescopes (SOFIA/HAWC+, JCMT/POL2) and present the results in followup papers.

\end{document}